\documentclass[oldversion]{aa}

\usepackage{graphicx}
\usepackage{graphics}
\usepackage{txfonts}
\usepackage{natbib}
\bibpunct{(}{)}{;}{a}{}{,} 

\begin{document}

\newcommand{\simgt}{\lower.5ex\hbox{$\;\buildrel>\over\sim\;$}}
\newcommand{\simlt}{\lower.5ex\hbox{$\;\buildrel<\over\sim\;$}}
\newcommand{\hst}{{\sl HST}}
\newcommand{\spitzer}{{\sl Spitzer}}
\newcommand{\lya}{{Ly$\alpha$}}                      
\newcommand{\bra}{{Br$\alpha$}}                      
\newcommand{\brg}{{Br$\gamma$}}                      
\newcommand{\ha}{{H$\alpha$}}                      
\newcommand{\hb}{{H$\beta$}}             
\newcommand{\zsun}{{$Z_\odot$}}                      
\newcommand{\msun}{{$M_\odot$}}                      
\newcommand{\lsun}{{$L_\odot$}}                      
\newcommand{\av}{{$A_V$}}
\newcommand{\vi}{{$V-I$}}
\newcommand{\hii}{{H{\sc ii}}}
\newcommand{\hi}{{H{\sc i}}}
\newcommand{\Ne}{{N$_e$}}
\newcommand{\magsq}{mag\,arcsec$^{-2}$}
\newcommand{\micron}{\,$\mu$m}

\newcommand{\nii}{\ensuremath{\mathrm{[N\,II]}}}
\newcommand{\oiii}{\ensuremath{\mathrm{[O\,III]}}}
\newcommand{\oii}{\ensuremath{\mathrm{[O\,II]}}}
\newcommand{\sii}{\ensuremath{\mathrm{[S\,II]}}}

\newcommand{\Ho}{\ensuremath{\mathrm{H}_0}}
\newcommand{\Msun}{\ensuremath{~\mathrm{M}_\odot}}
\newcommand{\Lsun}{\ensuremath{~\mathrm{L}_\odot}}
\newcommand{\LBsun}{\ensuremath{~\mathrm{L}_{B\odot}}}
\newcommand{\AV}{\ensuremath{\mathrm{A}_V}}
\newcommand{\QH}{\ensuremath{Q(\mathrm{H})}}
\newcommand{\bh}{\ensuremath{\mathrm{BH}}}
\newcommand{\mbh}{\ensuremath{M_\mathrm{BH}}}
\newcommand{\mdyn}{\ensuremath{M_\mathrm{dyn}}}
\newcommand{\lsph}{\ensuremath{L_\mathrm{bul}}}
\newcommand{\LV}{\ensuremath{L_\mathrm{V}}}
\newcommand{\lopt}{\ensuremath{L_\mathrm{opt}}}


%
\newcommand{\mlr}{\ensuremath{\Upsilon}}
\newcommand{\I}{\ensuremath{i}}
\newcommand{\Th}{\ensuremath{\theta}}
\newcommand{\B}{\ensuremath{b}}
\newcommand{\So}{\ensuremath{\mathrm{s}_\circ}}
\newcommand{\Vsys}{\ensuremath{V_\mathrm{sys}}}
\newcommand{\chisq}{\ensuremath{\chi^2}}
\newcommand{\chisqr}{\ensuremath{\chi^2_\mathrm{red}}}
\newcommand{\chisqc}{\ensuremath{\chi^2_\mathrm{c}}}
\newcommand{\nnn}{NGC\,3627}

\title{Molecular Gas in NUclei of GAlaxies (NUGA) \\ 
XIV. The barred LINER/Seyfert 2 galaxy NGC\,3627
\thanks{Based on observations
carried out with the IRAM Plateau de Bure Interferometer. IRAM is supported by the
INSU/CNRS (France), MPG (Germany), and IGN (Spain).}
}

\author{V. Casasola \inst{1,2},
L.K. Hunt \inst{1}, 
F. Combes \inst{3}, 
S. Garc\'ia-Burillo \inst{4},  
\and R. Neri \inst{5}
}
\offprints{{\tt casasola@ira.inaf.it}}

\institute{
INAF-Osservatorio Astrofisico di Arcetri, Largo E. Fermi, 5, 50125 Firenze, Italy \and
INAF-Istituto di Radioastronomia, via Gobetti 101, 40129 Bologna, Italy \and
Observatoire de Paris, LERMA, 61 Av. de l'Observatoire, F-75014, Paris, France 
\and Observatorio Astron\'omico Nacional (OAN) - Observatorio de Madrid, C/ Alfonso XII, 3, 28014 Madrid, Spain
\and IRAM-Institut de Radio Astronomie Millim\'etrique, 300 Rue de la Piscine,
38406-St.Mt.d`H\`eres, France
}

\date{Received ; accepted}

\abstract{
We present $^{12}$CO(1--0) and $^{12}$CO(2--1) maps of the 
interacting barred LINER/Seyfert 2 galaxy \nnn\ obtained 
with the IRAM interferometer at resolutions of 
2\farcs1 $\times$ 1\farcs3 and 0\farcs9 $\times$ 0\farcs6, 
respectively. 
We also present single-dish IRAM 30\,m $^{12}$CO(1--0) and 
$^{12}$CO(2--1) observations  
used to compute short spacings 
and complete interferometric measurements.
These observations are complemented by IRAM 30\,m measurements
of HCN(1--0) emission detected in the center of \nnn. 
The molecular gas emission shows a nuclear peak, an elongated bar-like 
structure of $\sim$18\arcsec\ ($\sim$900\,pc) diameter in both $^{12}$CO maps and, 
in $^{12}$CO(1--0), a two-arm spiral feature from $r$$\sim$9\arcsec\ ($\sim$450\,pc) 
to $r\sim$16\arcsec\ ($\sim$800\,pc).
The inner $\sim$18\arcsec\ bar-like structure, with a north/south 
orientation (PA = 14$^{\circ}$), forms two peaks at the extremes of this 
elongated emission region.
The kinematics of the inner molecular gas shows signatures of non-circular 
motions associated both with the 18\arcsec\ bar-like structure and the spiral 
feature detected beyond it.
The 1.6\,$\mu$m $H$-band 2MASS image of \nnn\ shows a stellar bar with a
PA = $-21^{\circ}$, different from the PA (= $14^{\circ}$)  
of the $^{12}$CO bar-like structure, indicating that 
the gas is leading the stellar bar.
The far-infrared \spitzer-MIPS 70 and 160\,$\mu$m images of \nnn\
show that the dust emission is intensified at the nucleus and 
at the ansae at the ends of the bar, coinciding with the $^{12}$CO peaks. 
The \textit{GALEX} far-ultraviolet (FUV) morphology of \nnn\ displays 
an inner elongated (north/south) ring delimiting a hole around 
the nucleus, and the $^{12}$CO bar-like structure is contained 
in the hole observed in the FUV.
The torques computed with the \hst-NICMOS F160W image and our PdBI
maps are negative down to the resolution limit of our images, $\sim$60 pc
in $^{12}$CO(2--1).
If the bar ends at $\sim$3\,kpc, 
coincident with corotation (CR), 
the torques are negative between the CR of the bar and the nucleus, 
down to the resolution limit of our observations.
This scenario is compatible with a recently-formed rapidly rotating bar which has had insufficient 
time to slow down because of secular evolution, and thus has not yet formed an inner
Lindblad resonance (ILR).
The presence of molecular gas inside the CR of the primary bar, 
where we expect that the ILR will form, makes \nnn\ a potential 
\textit{smoking gun} of inner gas inflow. 
The gas is fueling the central region, and in a second step 
could fuel directly the active nucleus.
\keywords{galaxies: individual: NGC~3627 -- galaxies: spiral -- galaxies: active --
galaxies: nuclei -- galaxies: ISM -- galaxies: kinematics and dynamics}
}

\authorrunning{Casasola et al.} 
\titlerunning{NUGA XIV: NGC\,3627}

\maketitle

\section{Introduction}
The Nuclei of Galaxies (NUGA) project \citep[][]{santi03} 
is an IRAM Plateau de Bure Interferometer (PdBI) and 30\,m single-dish 
survey of nearby low-luminosity active galactic nuclei (LLAGN).
The aim is to map, at high resolution ($\sim$0\farcs5-2\arcsec)
and high sensitivity ($\sim$2-4\,mJy\,beam$^{-1}$), the 
distribution and dynamics of the molecular gas in the inner kpc of the 
galaxies of our sample, and to study the different mechanisms for gas 
fueling of LLAGN.

NUGA galaxies analyzed so far show that there is no unique 
circumnuclear molecular gas feature linked with nuclear activity, but rather a 
variety of molecular gas morphologies which characterize the inner kpc of 
active galaxies.
We have found one- and two-armed instabilities \citep{santi03}, well-ordered 
rings and nuclear spirals \citep{francoise04,vivi08a}, circumnuclear 
asymmetries \citep{melanie05}, large-scale bars \citep{fred07,leslie08}, 
and smooth disks \citep{vivi10}.
Among these morphologies, analyzing the torques exerted by 
the stellar gravitational potential on the molecular gas shows that 
only four NUGA galaxies: 
NGC\,6574 \citep{lindt-krieg08},
NGC\,2782 \citep[][]{leslie08},
NGC\,3147 \citep{vivi08a}, 
and NGC\,4579 \citep[][]{santi09}
show evidence for gas inflow.
These galaxies have several features in common:
(1)~a large circumnuclear mass concentration (i.e., a dominant stellar bulge);
(2)~a high circumnuclear molecular gas fraction ($\ga$10\%); and
(3)~kinematically decoupled bars with overlapping dynamical resonances.
The large amount of gas around the nucleus, combined with dynamical features 
that enable the gas to penetrate the inner Lindblad Resonance (ILR),
seem to be necessary (and perhaps sufficient) ingredients for inducing
gas inflow in circumnuclear scales.

The existence of different nuclear molecular morphologies 
can be sought in the variety of timescales characterizing nuclear 
activity.
Strong fueling only lasts for a time of 
$t_{\rm fuel}\sim0.002\times t_{\rm H}$, where 
$t_{\rm H}\sim1.4\times10^{10}$\,yr is the age of the 
Universe \citep{heckman04}.
Thus, the total time during which strong fueling can occur
is around $t_{\rm fuel}\sim3\times10^{7}$\,yr; if
there are $N$ fueling events per black hole per Hubble time, 
each event would have a duration of 
$t_{\rm event}\sim3\times10^{7}/N$\,yr.
This implies that the strong accretion phase is a fraction
$\simeq 0.3/N$ of the characteristic galaxy dynamical time ($\sim10^{8}$\,yr).
Although large-scale bars can produce gas inflow  
\citep[e.g.,][]{francoise85,sakamoto99} and in some cases also drive 
powerful starbursts \citep[e.g.,][]{knapen02,jogee05}, a correlation 
between large-scale bars and nuclear activity has not yet been verified  
\citep[e.g.,][]{mulchaey97}.
This lack of correlation is probably related to the different 
timescales for bar-induced gas inflow \citep[$\gtrsim$300 Myr,][]{jogee05}, 
AGN duty cycles ($\sim$$10^{7}$\,yr), and intermittent active accretion 
every $\sim$$10^{8}$\,yr 
\citep[][]{ferrarese01,marecki03,janiuk04,hopkins06,king07}.
The comparison of these different timescales suggests that most AGN are 
in an intermediate phase between active accretion episodes making the detection 
of galaxies with nuclear accretion somewhat difficult.

Gravitational torques act on timescales of $\sim$10$^{6-7}$\,yr and 
are the most efficient mechanism in driving the gas from large 
spatial scales (some tens of kpc) to intermediate spatial scales 
(a few hundreds of pc).
Dynamical friction and viscous torques are often invoked, in addition 
to gravitational torques, as possible mechanisms of AGN fueling.
However, dynamical friction of giant molecular clouds in the stellar bulge of a galaxy
tends to be a slow, inefficient process which, to first approximation,
can be neglected relative to gravity torques \citep{santi05}.
Viscous torques can be more effective, and are favored in the presence of
large density gradients and high galactic shear \cite[see][for details]{santi05}.
Nevertheless, they are relatively inefficient when there are strong (positive)
gravity torques.

This paper is dedicated to the galaxy \nnn, the eleventh
object of the core NUGA sample studied on a case-by-case basis.
\nnn\ (Messier 66, $D$ = 10.2\,Mpc, $H_{0}$ = 73 km s$^{-1}$ Mpc$^{-1}$) 
is an interacting \citep[e.g.,][]{vivi04} and barred galaxy classified as SAB(s)b 
showing signatures of a LINER/Seyfert 2 type nuclear activity \citep{ho97}.
With NGC~3623 and NGC~3628, it forms the well-known Leo Triplet 
(M\,66 Group, VV\,308).
Since the discovery of a long plume in H{\scshape i} extending about 50$^{\prime}$ 
to the east of NGC~3628 \citep{zwicky56,haynes79}, evidence of past 
interactions between \nnn\ and NGC~3628 (the two largest 
spirals in the group), the Leo Triplet has been extensively studied from 
the radio to the optical, and in X-ray bandpasses. 
Optical broad-band images of \nnn\ reveal a pronounced and asymmetric 
spiral pattern with heavy dust lanes, indicating strong density wave 
action \citep{ptak06}.
While the western arm is accompanied by weak traces of 
star formation (SF) visible in H$\alpha$, the eastern arm contains a star-forming 
segment in its inner part \citep{smith94}.
\nnn\ also possesses X-ray properties of a galaxy with a recent
starburst \citep{dahlem96}.
Both the radio continuum (2.8\,cm and 20\,cm) and the $^{12}$CO(1--0)
emissions show a nuclear peak, extend along 
the leading edges of the bar forming two broad maxima 
at the bar ends, and then the spiral arms  trail off from 
the  bar ends \citep{haan08,paladino08,haan09}.
On the contrary, the H{\scshape i} emission exhibits a spiral morphology 
without signatures of a bar in the atomic gas \citep{haan08,walter08,haan09}. 

\begin{table}
\caption[]{Fundamental parameters for \nnn.}
\begin{center}
\begin{tabular}{lll}
\hline
\hline
Parameter  & Value$^{\mathrm{b}}$ & Reference$^{\mathrm{c}}$ \\
\hline
$\alpha_{\rm J2000}$$^{\mathrm{a}}$ & 11$^h$20$^m$15.02$^s$ & (1) \\
$\delta_{\rm J2000}$$^{\mathrm{a}}$ & 12$^{\circ}$59$^{\prime}$29\farcs50 & (1) \\
$V_{\rm hel}$ & 744 km\,s$^{-1}$ & (1) \\
RC3 Type & SAB(s)b & (2) \\
Nuclear Activity & LINER/Seyfert 2 & (3) \\
Inclination & 61\fdg3 & (1) \\
Position Angle & 178$^{\circ}$ $\pm$ 1$^{\circ}$ & (1) \\
Distance & 10.2\,Mpc (1\arcsec\ = 49\,pc) & (2) \\
L$_{B}$ & $4.2 \times 10^{10}$\,L$_{\odot}$ & (4) \\
M$_{\rm H\,I}$ & $8.1 \times 10^{8}$\,M$_{\odot}$ & (5) \\
M$_{\rm H_{2}}$ & $4.1 \times 10^{9}$\,M$_{\odot}$ & (6) \\
M$_{\rm dust}$(60 and 100\,$\mu$m)& $4.5 \times 10^{6}$\,M$_{\odot}$ & (4) \\
L$_{\rm FIR}$ & $1.2 \times 10^{10}$\,L$_{\odot}$ & (7) \\
\hline
\hline
\end{tabular}
\label{table1}
\end{center}
\begin{list}{}{}
\item[$^{\mathrm{a}}$] ($\alpha_{\rm J2000}$, $\delta_{\rm J2000}$) is the
phase tracking center of our $^{12}$CO interferometric observations, 
assumed coincident with the dynamical center of \nnn\ (see Sect. \ref{sec:dyncen}).
\item[$^{\mathrm{b}}$] 
Luminosity and mass values extracted from the literature 
have been scaled to the distance of $D$ = 10.2\,Mpc.
\item[$^{\mathrm{c}}$] 
(1) This paper;
(2) NASA/IPAC Extragalactic Database (NED, http://nedwww.ipac.caltech.edu/); 
(3) \citet{ho97}; 
(4) \citet{vivi04}; 
(5) \citet{haan08}; 
(6) \citet{kuno07}; 
(7) {\it IRAS} Catalog.
\end{list} 
\end{table}

\begin{table*}
\caption[]{$^{12}$CO(1--0) flux values, both obtained by our 
observations and extracted from the literature, for \nnn.}
\begin{center}
\begin{tabular}{llllll}
\hline
\hline
Reference & Telescope &	Diameter & Primary beam	or FOV$^{\mathrm{a}}$ & Beam & Flux \\
& & [m] &	[\arcsec] & [\arcsec $\times$ \arcsec] &	[Jy km\,s$^{-1}$] \\
\hline
\citet{young95}	 & FCRAO	   & 14 & 45 &		        & 786	\\
This paper	 & PdBI+30\,m &	& 42 & 2.1 $\times$ 1.3	& 668	\\
This paper	 & PdBI+30\,m & 	& 22$^{\mathrm{b}}$ & 2.1 $\times$ 1.3	& 359 \\
This paper	 & PdBI	   &	& 22$^{\mathrm{b}}$ & 2.0 $\times$ 1.3	& 251	\\
This paper	& 30\,m	   & 30	& 22 (central position) & & 343$^{\mathrm{c}}$	\\
\citet{helfer03} & NRAO    & 12 & 55 (inner 50\arcsec$\times$50\arcsec) 
&  & 1100--1200	\\
This paper      & 30\,m	   & 30	& 22 (inner 50\arcsec$\times$50\arcsec) 
& &1097$^{\mathrm{d}}$ \\
\hline
\hline
\end{tabular}
\label{table2}
\end{center}
\begin{list}{}{}
\item[$^{\mathrm{a}}$] 
Primary beam is considered for single-dish observations, while field-of-view (FOV) 
for interferometric or combined (interferometric+single-dish) ones.
\item[$^{\mathrm{b}}$] 
The photometry has been performed within 22\arcsec, the $^{12}$CO(1--0) primary 
beam for the 30\,m telescope.
\item[$^{\mathrm{c}}$] 
The $^{12}$CO(1--0) recovered flux for the central position (0\arcsec, 0\arcsec).
\item[$^{\mathrm{d}}$] 
The $^{12}$CO(1--0) recovered flux for inner 
$\sim$50\arcsec$\times$50\arcsec, $5 \times 5$ mapping with 7\arcsec 
spacing (see Sect. \ref{sec:30m-obs}).
\end{list} 
\end{table*}

The most recent \hi\ mass determination for \nnn\ has been obtained 
by \citet[][]{haan08}, M$_{\rm H\,I}=8.1\times$10$^{8}$\,M$_{\odot}$ 
(reported to our adopted distance of $D=10.2$\,Mpc), on average less than 
the typical value expected for interacting galaxies of the same 
Hubble type \citep[][]{vivi04}.
The H$_{2}$ mass content estimated by \citet{kuno07} is 
$4.1\times10^{9}\,$M$_{\odot}$ (scaled to our distance of 
$D=10.2$\,Mpc for \nnn).

These H$_{2}$ and \hi\ mass values give a H$_{2}$/\hi\ mass ratio of 5.1, 
high compared to the average ratio expected for 
galaxies similar to \nnn, M$_{\rm H_{2}}/$M$_{\rm HI} = 0.9$ 
\citep[][]{vivi04}.
The high H$_{2}$/\hi\ mass ratio in \nnn\ is probably due 
to the tidal interaction with NGC\,3628, since this galaxy has ``captured''
much of the \hi\ in \nnn\ \citep{zhang93}.

Other molecular transitions have been detected in \nnn, 
including HCN(1--0), HCN(2--1), HCN(3--2), HCO$^{+}$(1--0), and 
HCO$^{+}$(3--2), suggesting the presence of high density 
gas \citep[][]{gao04,melanie08}.
We list in Table \ref{table1} the main observational parameters
of \nnn.

The structure of this paper is as follows. In Sect. \ref{sec:obs}, 
we describe our new observations of \nnn\ and the literature data 
with which we compare them.
In Sects. \ref{sec:30m} and \ref{sec:pdbi}, we present the observational 
results, both single-dish and interferometric, describing morphology,
excitation conditions, and kinematics of the molecular gas in the inner kpc of
\nnn.
Comparisons between $^{12}$CO observations and those obtained at other
wavelengths are given in Sect. \ref{sec:comparison}.
In Sect. \ref{sec:torques}, we describe the computation of the gravity 
torques derived from the stellar potential in the inner region of \nnn, 
and in Sect. \ref{sec:interpretation}, we give an dynamical 
interpretation of the results.
Finally, Sect. \ref{sec:conclusions} summarizes our main results.

We assume a distance to \nnn\ of $D=10.2$\,Mpc,
(HyperLeda DataBase\footnote{\citet[][]{paturel03}, 
http://leda.univ-lyon1.fr}) and a Hubble constant 
$H_0=73$\,km\,s$^{-1}$\,Mpc$^{-1}$.
This distance means that 1\arcsec\ corresponds to 49\,pc.

\begin{figure*}
\centering
\begin{tabular}{c}
\includegraphics[height=0.8\textwidth,angle=-90]{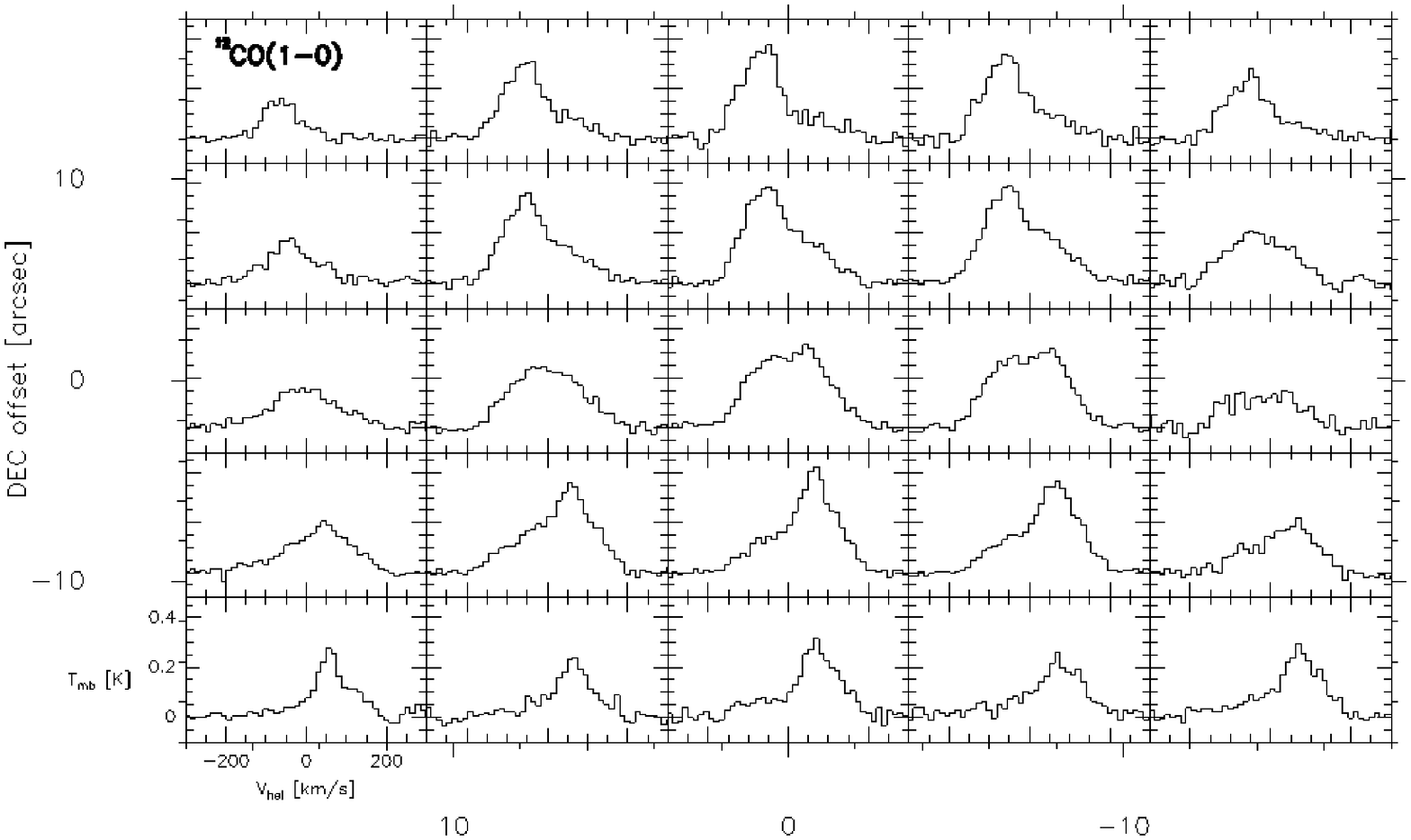}\\
\includegraphics[height=0.8\textwidth,angle=-90]{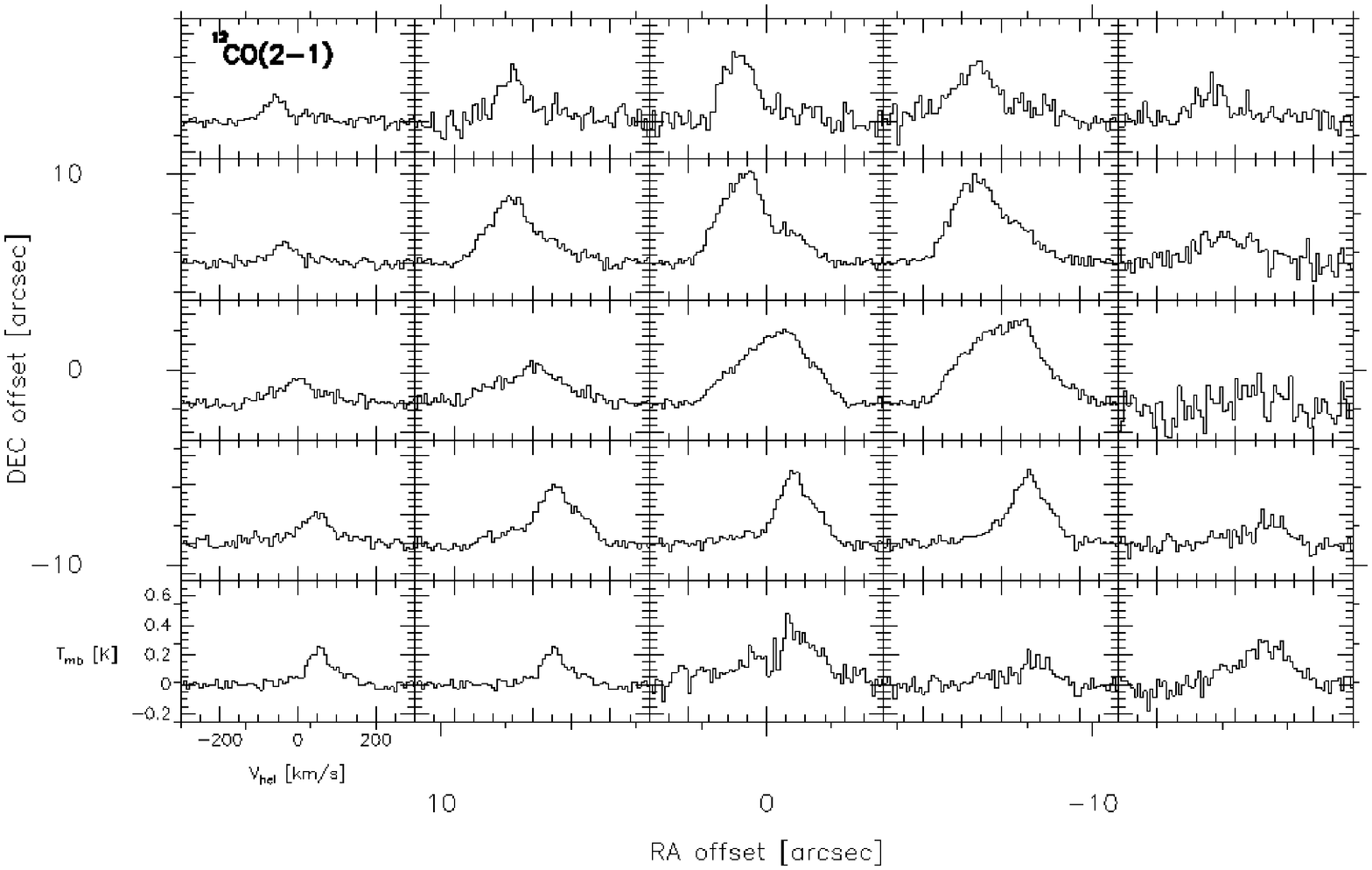}\\
\end{tabular}
\caption{
Spectra maps of \nnn\ made with the IRAM 30\,m with
7\arcsec\ spacing in $^{12}$CO(1--0) [top] and $^{12}$CO(2--1) [bottom]. 
The positions are arcsec offsets relative to the phase tracking center of our
interferometric observations (see Table \ref{table1}).
Each spectrum has a velocity scale from
$-300$ to $300\,{\rm km\,s^{-1}}$, and a beam-averaged radiation temperature
scale (T$_{\rm mb}$) from $-0.10$ to $0.48\,{\rm K}$ for $^{12}$CO(1--0) and 
from $-0.25$ to 0.70\,K for $^{12}$CO(2--1).
}
\label{fig:n3627-30m}
\end{figure*}

\section{Observations \label{sec:obs}}
\subsection{Interferometric observations \label{sec:PdB-obs}}
We observed \nnn\ with the IRAM PdBI (6 antennas) in the ABCD 
configuration of the array between 2003 September and 2004 February  
in the $^{12}$CO(1--0) [115\,GHz] and the $^{12}$CO(2--1) [230\,GHz]
line.
The PdBI receiver characteristics, the observing procedures, and the 
image reconstruction are similar to those described in \citet{santi03}.
The quasar 3C454.3 was used for bandpass calibration, 
3C273 for flux calibration, and 1546+027 for phase and amplitude 
calibrations.

Data cubes with 512 $\times$ 512 pixels (0\,\farcs27\,${\rm pixel}^{-1}$
for $^{12}$CO(1--0) and 0\,\farcs13\,${\rm pixel}^{-1}$ for $^{12}$CO(2--1))
were created over a velocity interval of -242.5 km\,s$^{-1}$ to
+242.5 km\,s$^{-1}$ in bins of 5 km\,s$^{-1}$.
The images presented here were reconstructed using the standard
IRAM/GILDAS\footnote{http://www.iram.fr/IRAMFR/GILDAS/}
software \citep{guilloteau} and restored with Gaussian beams of 
dimensions 2\farcs0 $\times$ 1\farcs3 (PA = 23$^{\circ}$) at 115\,GHz
and 0\farcs9 $\times$ 0\farcs6 (PA = 28$^{\circ}$) at 230\,GHz.
We used natural and uniform weightings to generate
$^{12}$CO(1--0) and $^{12}$CO(2--1) maps, respectively.
This allows to maximize the flux recovered in $^{12}$CO(1--0)
and optimize the spatial resolution in $^{12}$CO(2--1).
In the cleaned maps, the {\it rms} levels are 
$3.7\,{\rm mJy\,beam^{-1}}$ and $6.7\,{\rm mJy\,beam^{-1}}$ 
for the $^{12}$CO(1--0) and $^{12}$CO(2--1) lines,
respectively at a velocity resolution of 5\,km\,s$^{-1}$. 
At a level of 3$\sigma$ no 3\,mm (1\,mm) continuum was detected toward
\nnn\, down to an {\it rms} noise level of $0.34\,{\rm mJy\,beam^{-1}}$ 
($0.48\,{\rm mJy\,beam^{-1}}$).
The conversion factors between intensity and brightness temperature
are $34\,{\rm K\,(Jy\,beam^{-1})^{-1}}$ at 115\,GHz and
$41\,{\rm K\,(Jy\,beam^{-1})^{-1}}$ at 230\,GHz.
All velocities are referred to the systemic velocity
$V_{\rm sys, hel}$ = 744 km\,s$^{-1}$ and 
$(\Delta \alpha, \Delta \delta)$ offsets are relative to the phase 
tracking center of our observations 
(11$^h$20$^m$15.02$^s$, 12$^{\circ}$59$^{\prime}$29.50\arcsec)
[see later Sect. \ref{sec:dyncen}].
All maps are centered on this position (see Table \ref{table1}) and are 
not corrected for primary beam attenuation.

\subsection{Single-dish observations \label{sec:30m-obs}}
We performed IRAM 30\,m telescope observations of \nnn\ 
on July 16-19, 2002, in a $5 \times 5$ raster pattern with 7\arcsec\ 
spacing.
By using 4 SIS receivers, we simultaneously observed the frequencies 
of the $^{12}$CO(1--0) [115\,GHz], the $^{12}$CO(2--1) [230\,GHz], 
and the HCN(1--0) [89\,GHz] lines.
The $^{12}$CO(2--1) line has been observed in dual-polarization.
The half power beam widths (HPBW) are 22\arcsec, 12\arcsec, and 
29\arcsec\ for $^{12}$CO(1--0), $^{12}$CO(2--1), and HCN(1--0) lines, 
respectively.
Typical system temperatures were $\sim$110-145\,K at 115\,GHz,
$\sim$320-750\,K at 230\,GHz, and $\sim$110-145\,K at 89\,GHz.
For the single-dish data reduction, the Continuum and Line Analysis
Single-dish Software (CLASS$^{2}$) was used.
Throughout the paper we express the line intensity scale in units 
of the beam-averaged radiation temperature (T$_{\rm mb}$). 
T$_{\rm mb}$ is related to the equivalent antenna temperature
reported above the atmosphere (T$^{*}_{A}$) by 
$\eta=$T$^{*}_{\rm A}/$T$_{\rm mb}$, where $\eta$ is the telescope 
main-beam efficiency.
At 115\,GHz $\eta$ = 0.79, at 230\,GHz $\eta$ = 0.54, and at 89\,GHz 
$\eta$ = 0.82.
All observations were performed in ``wobbler-switching'' mode, with a minimum
phase time for spectral line observations of 2\,s and a maximum beam throw 
of 240\arcsec. The pointing accuracy was 
$\sim$3\arcsec\ {\it rms}.
The single-dish maps presented in this paper are centered on the phase tracking 
center of our interferometric observations (see Table \ref{table1}).

\subsection{Short spacing correction \label{sec:short-s-obs}}
An interferometer is limited by the minimum spacing of its antennas. 
Because two antennas can not be placed closer than some minimum 
distance ($D_{\rm min}$), signals on spatial scales larger than some size 
($\propto$$\lambda/D_{\rm min}$) will be attenuated. 
This effect, called the ``missing flux'' problem, is resolved by using 
single-dish observations to compute short spacings and complete the 
interferometric measurements.

By combining 30\,m and PdBI data, we found the best compromise between 
good angular resolution and complete restoration of the missing 
extended flux by varying the relative weights of 30\,m 
and PdBI observations.
The  combined  PdBI+30\,m maps have angular resolutions of 
2\farcs1 $\times$ 1\farcs3 at PA = $23^{\circ}$ for the 
$^{12}$CO(1--0) and 0\farcs9 $\times$ 0\farcs6 at 
PA = $30^{\circ}$ for the $^{12}$CO(2--1).
In the combined maps, the {\it rms} uncertainty $\sigma$
in 5\,km\,s$^{-1}$ width velocity channels is 3.6\,mJy\,beam$^{-1}$
and 6.5 \,mJy\,beam$^{-1}$ for the $^{12}$CO(1--0) and
$^{12}$CO(2--1) lines, respectively.
For these maps, the conversion factors between intensity and brightness 
temperature are $32\,{\rm K\,(Jy\,beam^{-1})^{-1}}$ at 115\,GHz and
$41\,{\rm K\,(Jy\,beam^{-1})^{-1}}$ at 230\,GHz.
All interferometric figures presented in this paper are realized 
with short-spacing-corrected data.

\begin{figure}
\centering
\includegraphics[height=8cm,angle=-90]{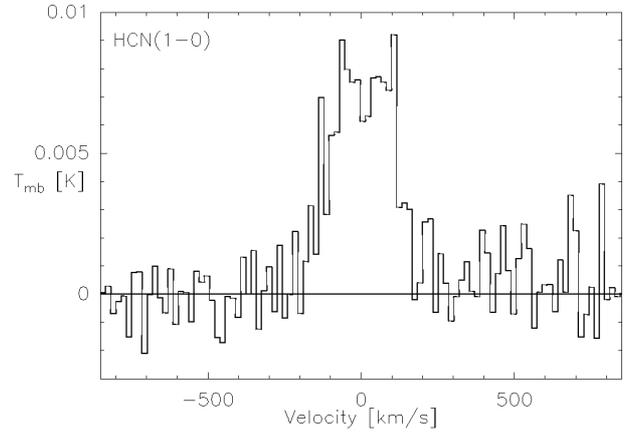}
\caption{HCN(1--0) spectrum toward the center of \nnn, averaged over
the 25-point map made with the IRAM 30\,m with 7\arcsec\ spacing.
The spectrum has a velocity scale from $-850$ to $850\,{\rm km\,s^{-1}}$ 
and a beam-averaged radiation temperature scale (T$_{\rm mb}$) from 
$-0.003$ to $0.010\,{\rm K}$.}
\label{fig:n3627-hcn}
\end{figure}

Within 22\arcsec, the $^{12}$CO(1--0) HPBW for the 30\,m telescope,
the map including only PdBI observations recovers a flux 
S$_{\rm CO(1-0)}$ = 251 Jy km s$^{-1}$, 70$\%$ of the total flux measured 
with the merged PdBI+30\,m map, S$_{\rm CO(1-0)}$ = 359 Jy km s$^{-1}$.
Table \ref{table2} reports both $^{12}$CO(1--0) flux values determined with 
our observations (single-dish, interferometric, and combined PdBI+30\,m) 
and those present in literature.
In this table, Col. (1) indicates the reference, Cols. (2) and (3) 
are the telescope (single-dish or interferometer) and the diameter 
of the single-dish telescope respectively, Col. (4) is the primary beam 
of the instrument or the diameter used for the performed photometry, 
Col. (5) is the beam in interferometric measurements, and 
Col. (6) gives the measured flux.
Table \ref{table2} shows that $^{12}$CO(1--0) fluxes we obtained with 
interferometric observations, single-dish, and combined measurements 
(PdBI+30\,m) are in good mutual agreement with each other and with data 
present in literature.
Our 30\,m observations give a value S$_{\rm CO(1-0)}$ = 343\,Jy\,km\,s$^{-1}$ 
for the central position, consistent with flux value found 
with PdBI+30\,m data within the $^{12}$CO(1--0) 30\,m--HPBW (22\arcsec).
The whole region covered with 30\,m observations 
($\sim$50\arcsec$\times$50\arcsec) gives a 
flux of 1097\,Jy km s$^{-1}$, in agreement with the BIMA SONG 
survey\footnote{Berkley-Illinois-Maryland Association Survey of 
Nearby Galaxies.} (NRAO 12\,m) measurements 
\citep[][see Fig. 50, $\sim$1100--1200\,Jy km s$^{-1}$]{helfer03}.
Moreover, within the 42\arcsec primary beam field of the PdBI, 
we recovered $\sim$85$\%$ of the flux detected by
\citet{young95} for the central position with the FCRAO 
(786\,Jy\,km\,s$^{-1}$), a good agreement considering
the uncertainties in the amplitude calibration and the non-correction 
by the primary beam attenuation.

\begin{figure*}
\centering
\hbox{
\includegraphics[width=0.3\textwidth,angle=-90]{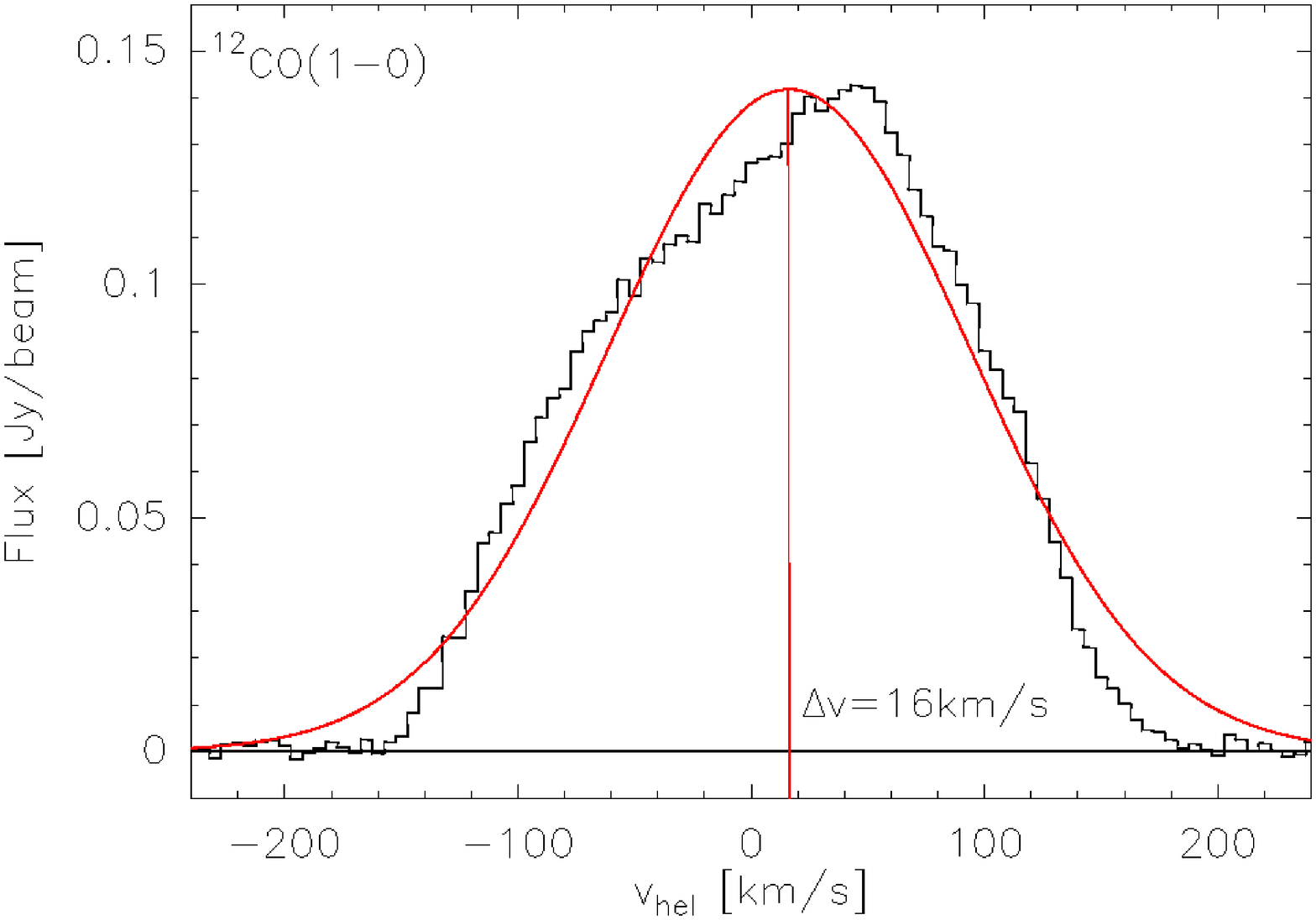}
\hspace{0.1\textwidth}
\includegraphics[width=0.3\textwidth,angle=-90]{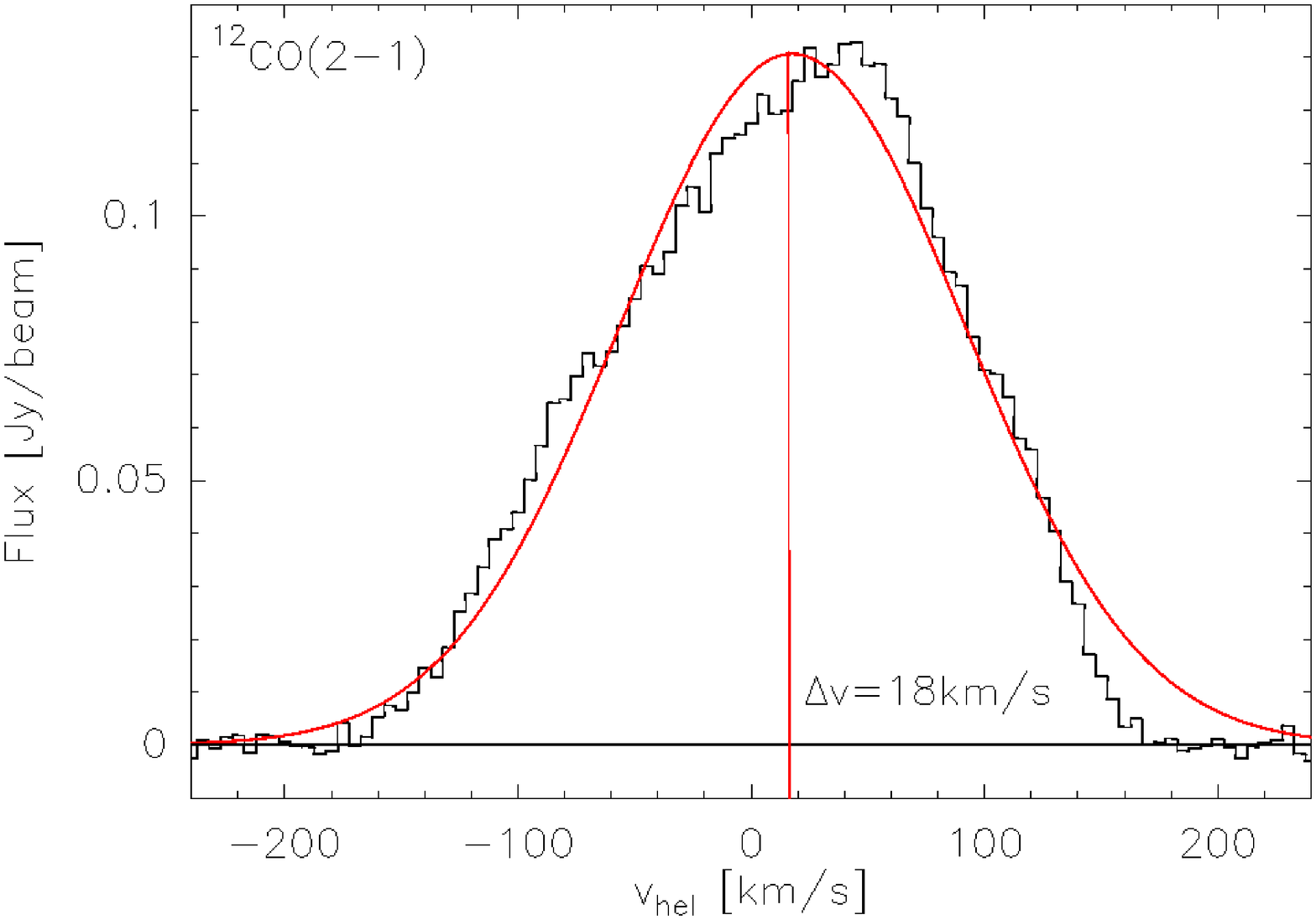}
}
\caption{
\textit{Left panel}: $^{12}$CO(1--0) integrated spectrum and 
gaussian fit (red) in the inner $\sim$2\arcsec\ of 
\nnn\ for PdBI+30\,m combined data. The gaussian fit shows that the 
heliocentric systematic velocity is redshifted by 16 km\,s$^{-1}$ with 
respect to the heliocentric velocity of the center (0 km\,s$^{-1}$).
\textit{Right panel}: Same for $^{12}$CO(2--1). The gaussian fit shows that
the heliocentric systematic velocity is redshifted by 18 km\,s$^{-1}$.
\label{fig:velhel}
}
\end{figure*}

\begin{figure*}
\centering
\includegraphics[width=0.43\textwidth,angle=-90]{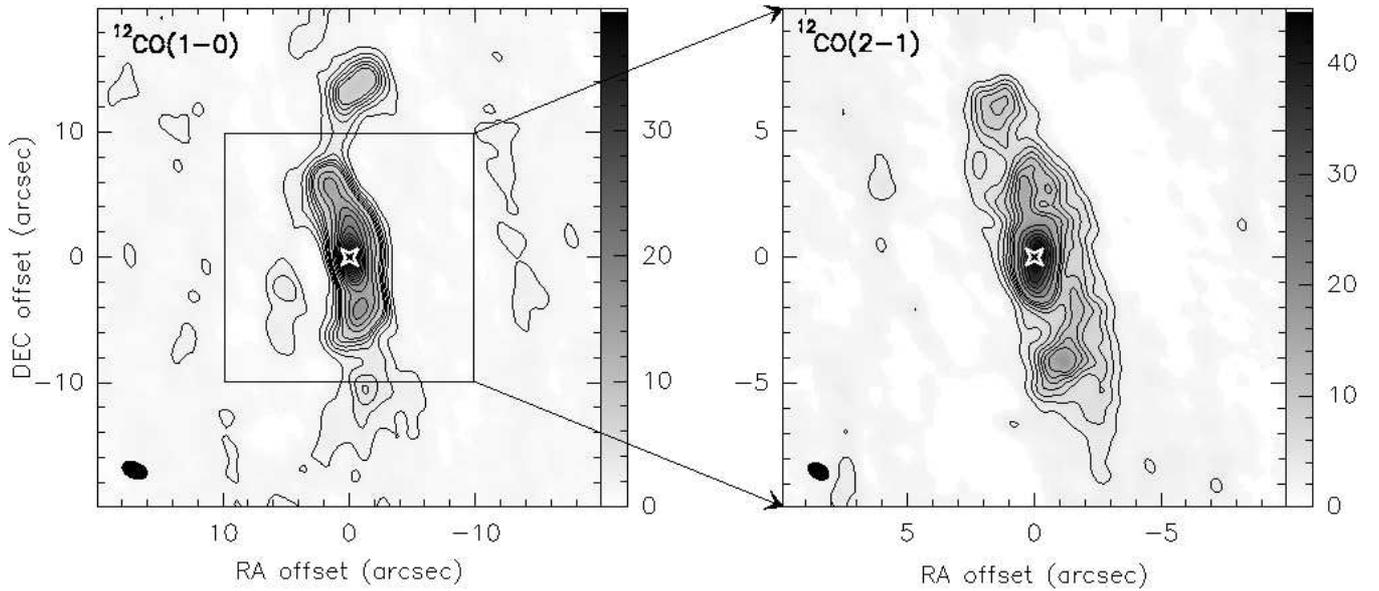}
\caption{\textit{Left panel}: $^{12}$CO(1--0) integrated intensity contours 
observed with the IRAM PdBI+30\,m toward the center of \nnn. 
The white star marks the coordinates 
of the dynamical center of the galaxy coincident with our phase tracking center 
(see Table \ref{table1}), with offsets in arcseconds.
The map, derived with 2$\sigma$ clipping, has not been corrected for
primary beam attenuation.
The {\it rms} noise level is $\sigma = 0.16\,{\rm Jy\,beam^{-1}\,km\,s^{-1}}$ 
and contour levels run from 3$\sigma$ to 33$\sigma$ with 
6$\sigma$ spacing and from 39$\sigma$ to the maximum with 
18$\sigma$ spacing.
In this map the $\pm 200\,{\rm km\,s^{-1}}$ velocity range is used. 
The beam of 2\farcs1 $\times$ 1\farcs3 is plotted in the lower left.
\textit{Right panel}: Same for $^{12}$CO(2--1).
The {\it rms} noise level is $\sigma = 0.30\,{\rm Jy\,beam^{-1}\,km\,s^{-1}}$ 
and contour levels run from 3$\sigma$ to 39$\sigma$ with 
6$\sigma$ spacing and from 45$\sigma$ to the maximum with 
18$\sigma$ spacing.
The beam of 0\farcs9 $\times$ 0\farcs6 is plotted at lower left.}
\label{fig:co10-21}
\end{figure*}

\subsection{Other images of \nnn\label{sec:otherdata}}
We also acquired the large-scale $^{12}$CO(1--0) emission 
image available thanks to the BIMA SONG 
survey performed with the 10-element BIMA millimeter 
interferometer \citep{welch96} at Hat Creek, California.
This image was first published by  \citet{regan01} and 
\citet{helfer03}, and covers a field of 350\arcsec$\times$410\arcsec 
(centered on the galaxy) with a pixel size of 1\arcsec, and a beam 
of 6\farcs6$\times$5\farcs5.

Several infrared (IR) images are included in our analysis: 
the \spitzer-IRAC 3.6\,$\mu$m image (to trace the stellar component), 
the \spitzer-IRAC 8\,$\mu$m image (to visualize the Polycyclic Aromatic 
Hydrocarbons [PAH] features), and the \spitzer-MIPS 70 and 160\,$\mu$m images 
(to study the dust emission and resolve the SF regions).
These IR images are available thanks to the project 
``SINGS: The Spitzer Infrared Nearby Galaxies Survey'' 
\citep{kennicutt03}.
The IRAC images cover a sky area of $\sim$1600\arcsec$\times$1890\arcsec 
and $\sim$1220\arcsec$\times$1420\arcsec at 3.6\,$\mu$m and  8\,$\mu$m respectively, 
both with a pixel size of 0\farcs75, 
and spatial resolutions of $\sim$1-2\arcsec.
The MIPS 70\,$\mu$m image 
covers $\sim$1940\arcsec$\times$3645\arcsec with a pixel size of 4\farcs5, 
and the MIPS 160\,$\mu$m image $\sim$2025\arcsec$\times$3460\arcsec with 
a pixel size of 9\arcsec.

We also use two 
near-infrared (NIR) images $H$ (1.65\,$\mu$m):
the first was taken from the Two Micron All Sky Survey (2MASS)
and covers a FOV of  $\sim$12$^{\prime}\times$12$^{\prime}$, 
with a resolution of 2\farcs5. 
The second 1.6\,$\mu$m $H$-band image of \nnn\ is available thanks to the F160W 
filter on the Near-Infrared Camera and Multi-Object Spectrometer 
(NICMOS, camera 3 [NIC3]) on board the Hubble Space Telescope (\hst). 
This image covers a FOV of 51\arcsec$\times$51\arcsec, has a resolution 
of 0\farcs2, and is not exactly centered on the galaxy but offset from our 
phase tracking center 7\arcsec\ toward west and 7\farcs4 toward south.
It is part of a survey of 94 nearby galaxies from the Revised 
Shapley Ames Catalog \citep[][]{boker99}.

Finally, we also use a far-ultraviolet (FUV) image from the \textit{GALEX}
satellite, whose band is centered at $\lambda_{eff}$ = 1516 \AA. 
This image has been already used and studied in the context of the 
\textit{GALEX} Nearby Galaxies Survey \citep[NGS,][]{gildepaz07}.
The image
covers a square region on the sky of size $\sim$5760\arcsec$\times$5760\arcsec, 
i.e., much larger than the extent of the optical disk of \nnn,
with 1\farcs5 pixels. 
As the image was reduced with the \textit{GALEX} data pipeline, it is already 
expressed in intensity units and skysubtracted.
The total FUV calibrated magnitude is 16.34$\pm$0.02,
corresponding to a FUV flux density of 1057$\pm$19 $\mu$Jy.

\begin{figure}
\centering
\includegraphics[width=0.4\textwidth,angle=-90]{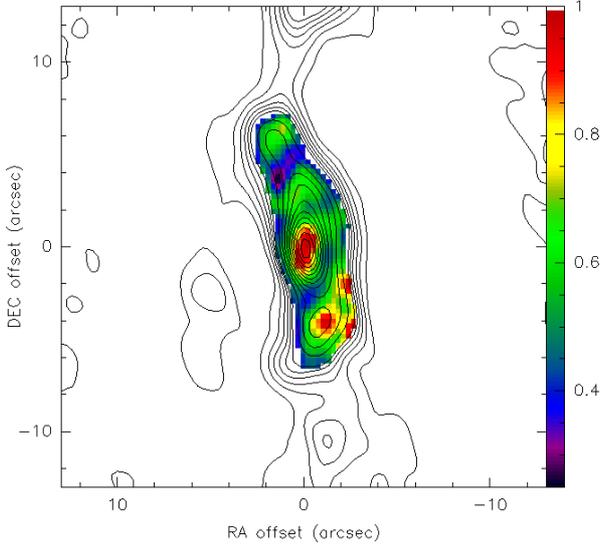}
\caption{
Color scale of the CO(2--1)/CO(1--0) ratio map and $^{12}$CO(1--0) intensity 
map contours as in Fig. \ref{fig:co10-21} 
(left panel).
}
\label{fig:ratio}
\end{figure}

\section{Single-dish results\label{sec:30m}}
The observations performed with the A and B receivers of the IRAM 30\,m 
telescope in the two $^{12}$CO lines covered the
inner $\sim$50\arcsec, corresponding to the central $\sim$2.5 kpc
(in diameter) of the galaxy (Fig. \ref{fig:n3627-30m}). 
The observed positions show that the central region of \nnn\ hosts 
extended molecular emission both in $^{12}$CO(1--0) and $^{12}$CO(2--1). 
The maximum detected T$_{\rm mb}$ is 0.4\,K in $^{12}$CO(1--0) at the 
offset position (0\arcsec, -7\arcsec), 
and 0.6\,K in $^{12}$CO(2--1) at the offset position 
(0\arcsec, 7\arcsec).

We estimate a flux of 1097 Jy\,km\,s$^{-1}$ within the inner 
$\sim$50\arcsec$\times$50\arcsec (see Table \ref{table2} and Sect. \ref{sec:short-s-obs}),  
in good agreement with previous single-dish flux determinations 
\citep[e.g.,][]{helfer03}.
Assuming a H$_{2}$-CO conversion factor of 
$X = N(\rm{H_{2}})$/$I\rm_{CO} = 2.2 \times 10^{20}$ cm$^{-2}$ 
(K km s$^{-1}$)$^{-1}$ \citep{solomon91}, the $^{12}$CO(1--0) integrated 
flux 
allows us to derive the H$_{2}$ mass within the observed region as:
\begin{eqnarray}M\rm_{H_{2}}\mbox{[M$_{\odot}$]} &=& 8.653 \times 10^{3}\,D^{2}\mbox{[Mpc]}\,S\rm_{CO(1-0)}\mbox{[Jy\,km\,s$^{-1}$]}
\label{h2mass}
\end{eqnarray}

\noindent
We derive an H$_{2}$ mass of 
M$\rm_{H_{2}}$$\sim$9.9$\times 10^{8} \rm{M_{\odot}}$ within
the inner $\sim$50\arcsec$\times$50\arcsec, and
taking into account the mass of helium, the total molecular 
mass is M$\rm_{mol} = M\rm_{H_{2}+He} = 1.36 \times M\rm_{H_{2}}$$
\sim$1.3$\times 10^{9} \rm{M_{\odot}}$.

The HCN(1--0) line has been observed for inner 25 positions 
with 7\arcsec\ spacing, covering the central $\sim$56\arcsec\ 
($\sim$2.7\,kpc in diameter).
The HCN(1--0) average spectrum over the 5$\times$5 grid is displayed 
in Fig. \ref{fig:n3627-hcn} and shows a peak at 
T$_{\rm mb}$$\sim$0.009\,K.
The HCN(1--0) velocity integrated intensity of the 
central position (0\arcsec, 0\arcsec) is 
I$_{\rm HCN(1-0)}$ = 3.1$\pm$0.3\,K\,km\,s$^{-1}$ with 
$\Delta$v = 237$\pm$32\,km\,s$^{-1}$, consistent with the results 
obtained by \citet{melanie08} 
for the same position observed
with the same instrument (I$_{\rm HCN(1-0)}$ = 2.7$\pm$0.2\,K\,km\,s$^{-1}$ with 
$\Delta$v = 290$\pm$30\,km\,s$^{-1}$).
The CO(1--0)/HCN(1--0) ratio averaged on the center of galaxy is roughly 
10, a value intermediate between the ratios found in spatially resolved 
molecular disks around AGN, such as NGC\,6951 \citep[][]{melanie07} and 
NGC\,1068 \citep[][]{melanie08}, and those found in pure starburst galaxies
such as M\,82 \citep[][]{melanie08}.

\begin{figure*}
\centering
\includegraphics[width=0.9\textwidth,angle=-90]{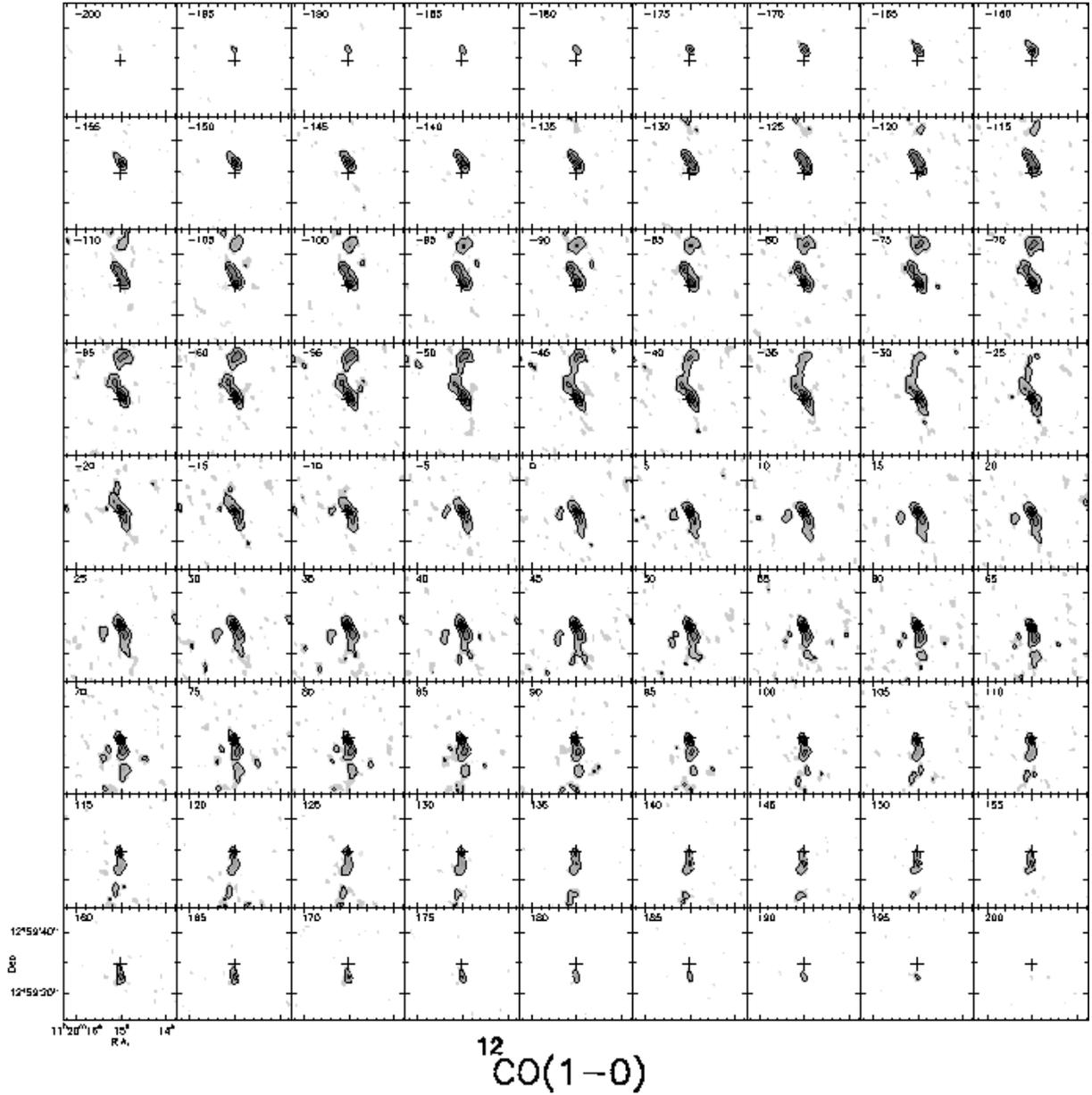}
\caption{$^{12}$CO(1--0) velocity channel maps observed with the 
IRAM PdBI+30\,m in the nucleus of \nnn, with a spatial resolution 
of 2\farcs1 $\times$ 1\farcs3 (HPBW).
The maps are centered on the phase tracking center of our observations 
($\alpha_{\rm J2000}$ = 11$^h$20$^m$15.02$^s$, $\delta_{\rm J2000}$ = 
12$^{\circ}$59$^{\prime}$29\farcs50) assumed 
to be coincident with the dynamical center of the galaxy.
Velocity channels range from $\Delta V = -200\,{\rm km\,s^{-1}}$ 
to $+200\,{\rm km\,s^{-1}}$ in steps of $5\,{\rm km\,s^{-1}}$ relative 
to $V_{\rm sys, hel}$ = 744 km\,s$^{-1}$ (see Sect. \ref{sec:dyncen}).
The contours run from 
$-40\,{\rm mJy\,beam^{-1}}$ to $260\,{\rm mJy\,beam^{-1}}$ 
with spacings of $60\,{\rm mJy\,beam^{-1}}$.}
\label{channels10}
\end{figure*}

\begin{figure*}
\centering
\includegraphics[width=0.9\textwidth,angle=-90]{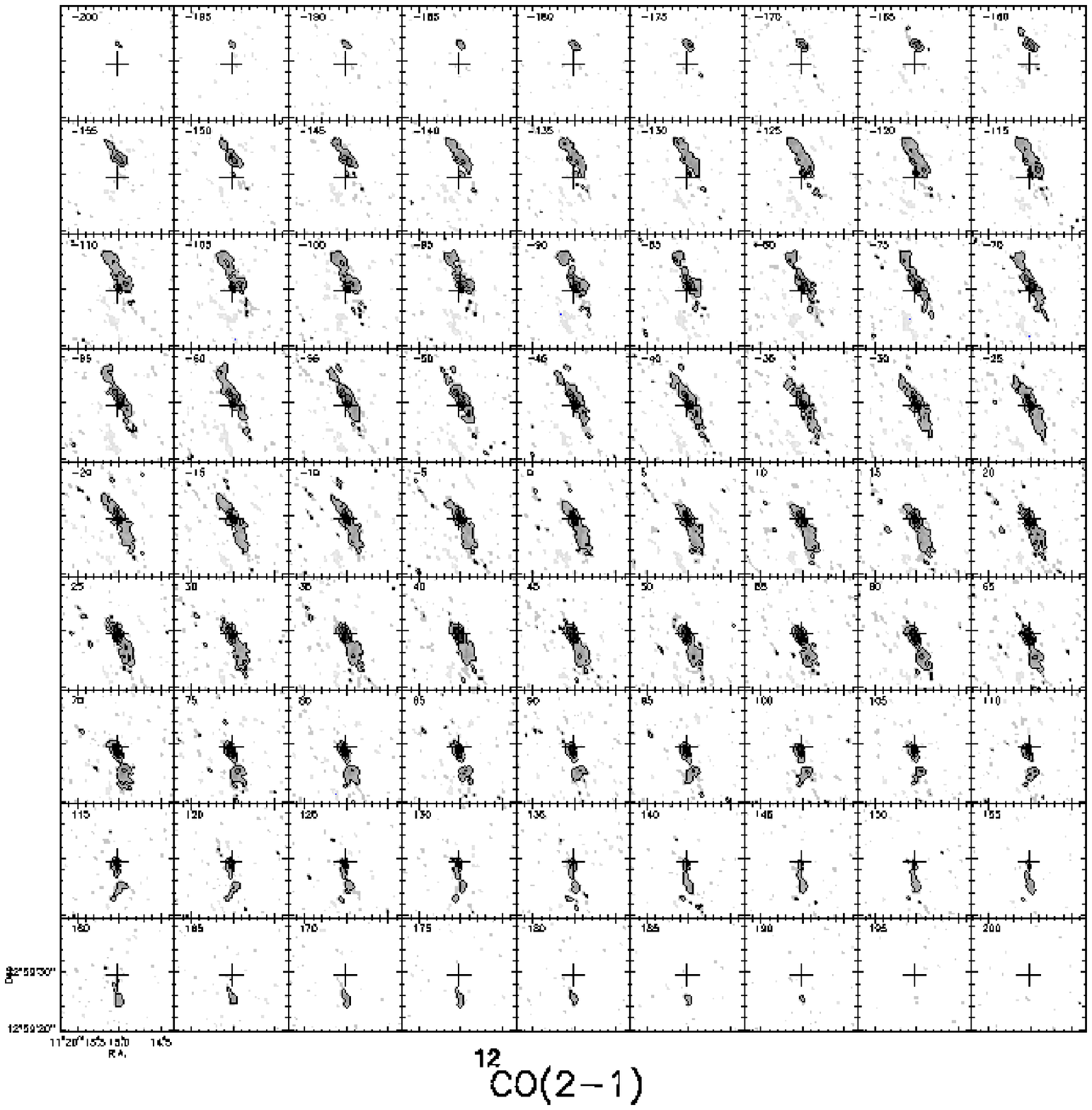}
\caption{Same as Fig. \ref{channels10} but for the $^{12}$CO(2--1) line, 
with a spatial resolution of 0\farcs9 $\times$ 0\farcs6.
The contours run from $-50\,{\rm mJy\,beam^{-1}}$ to 
$350\,{\rm mJy\,beam^{-1}}$ with spacings of $50\,{\rm mJy\,beam^{-1}}$.}
\label{channels21}
\end{figure*}

\section{Interferometric results\label{sec:pdbi}}
\subsection{Dynamical center\label{sec:dyncen}}
The phase tracking center of our observations
(see Table \ref{table1}) coincides almost exactly
with the nuclear radio source detected at 15\,GHz (VLA/2\,cm) by \citet{nagar00}
[11$^h$20$^m$15.01$^s$, 12$^{\circ}$59$^{\prime}$29\farcs76] 
and at 8.4\,GHz (VLA/3.6\,cm) by \citet{filho00} 
[11$^h$20$^m$15.0$^s$, 12$^{\circ}$59$^{\prime}$30\arcsec].
Thus, in the following, we assume that our observations are centered
on the dynamical center of \nnn.

The spectral correlators were centered at 114.992\,GHz 
for $^{12}$CO(1--0) and 229.979\,GHz for $^{12}$CO(2--1), 
corresponding to $V_{\rm LSR}$ = 727\,km\,s$^{-1}$.
Since for \nnn\ the difference between LSR and heliocentric velocity 
is $\sim$0 km\,s$^{-1}$, our observations were centered on 
$V_{\rm LSR}$ = $V_{\rm hel}\rm (PdBI)$ = 727 km\,s$^{-1}$.
In the inner $\sim$2\arcsec\ of \nnn\ the velocity centroid  
is 16 km\,s$^{-1}$ redshifted with respect to the heliocentric 
velocity of the center of our $^{12}$CO(1--0) observations 
(Fig. \ref{fig:velhel}, left panel).
Similarly to $^{12}$CO(1--0), for $^{12}$CO(2--1) we find that 
the velocity centroid is 18 km\,s$^{-1}$ redshifted with respect 
to the heliocentric velocity (Fig. \ref{fig:velhel}, right panel). 
Assuming an intermediate value  between the systemic heliocentric 
velocity determined for the $^{12}$CO(1--0) and that for the 
$^{12}$CO(2--1), we estimate $V_{\rm sys, hel}$ = 744 km\,s$^{-1}$.

This estimation of the systemic heliocentric velocity 
is 24 km\,s$^{-1}$ redshifted with respect to the
systemic heliocentric velocity determined from \hi\ observations 
\citep[720 km\,s$^{-1}$, HyperLeda Database;][]{haan08}.
In interacting galaxies and in those with a lopsided \hi\ morphology, 
a discrepancy between systemic velocity derived from $^{12}$CO and \hi\ 
observations is not unusual.
NGC\,4579 \citep{santi09} and  NGC\,5953 \citep{vivi10} exhibit 
differences of $\sim$50 km\,s$^{-1}$ between $^{12}$CO and 
\hi\ velocities, perhaps due to the interaction history of the 
galaxy and the different effect of the ram pressure on the atomic 
and molecular gas \citep[][]{santi09}.
In \nnn, the role of interaction history and the ram-pressure,
although not negligible, could have shifted the \hi\ barycenter with 
respect to the molecular one less strongly than in NGC\,4579  
and NGC\,5953.

\subsection{CO distribution and mass\label{sec:codistribution}}
Figure \ref{fig:co10-21} shows the $^{12}$CO(1--0) and $^{12}$CO(2--1) integrated 
intensity distributions in the inner $\sim$40\arcsec\ ($\sim$2\,kpc) of \nnn.
The $^{12}$CO(1--0) emission (Fig. \ref{fig:co10-21}, left panel) 
exhibits a peak at the nucleus, extends along a bar-like structure
of $\sim$18\arcsec\ ($\sim$900\,pc) diameter with a north/south 
orientation (see later Sect. \ref{sec:cokinematics} for the discussion on the PA)
and two peaks at its extremes, at $r$$\sim$5-6\arcsec\ ($\sim$270\,pc), 
with the southern one more evident.
The $^{12}$CO(1--0) morphology also shows a two-arm spiral feature from 
$r$$\sim$9\arcsec\ ($\sim$450\,pc) to $r$$\sim$16\arcsec\ ($\sim$800\,pc), with 
two peaks over these spiral arms at $r$$\sim$12-14\arcsec\ ($\sim$650\,pc).
A similar and more resolved distribution is found in $^{12}$CO(2--1) 
[Fig. \ref{fig:co10-21}, right panel]. 
Like the nuclear peak, the two peaks at the ends of the inner
$\sim$18\arcsec\ bar-like structure at $r$$\sim$5-6\arcsec, are more evident 
than in $^{12}$CO(1--0).

\begin{figure*}
\centering
\hbox{
\includegraphics[width=0.4\textwidth,angle=-90]{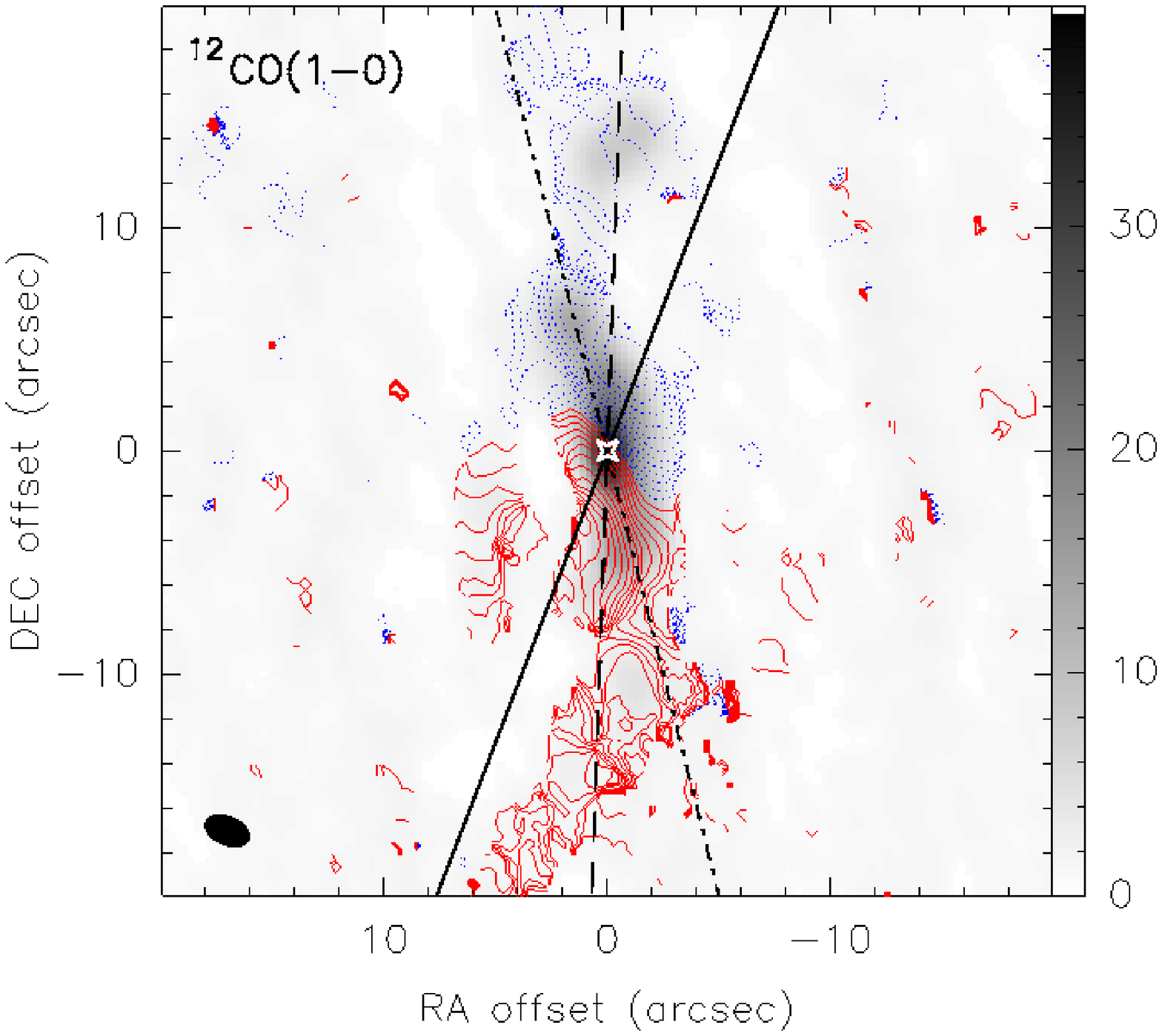}
\hspace{0.1\textwidth}
\includegraphics[width=0.4\textwidth,angle=-90]{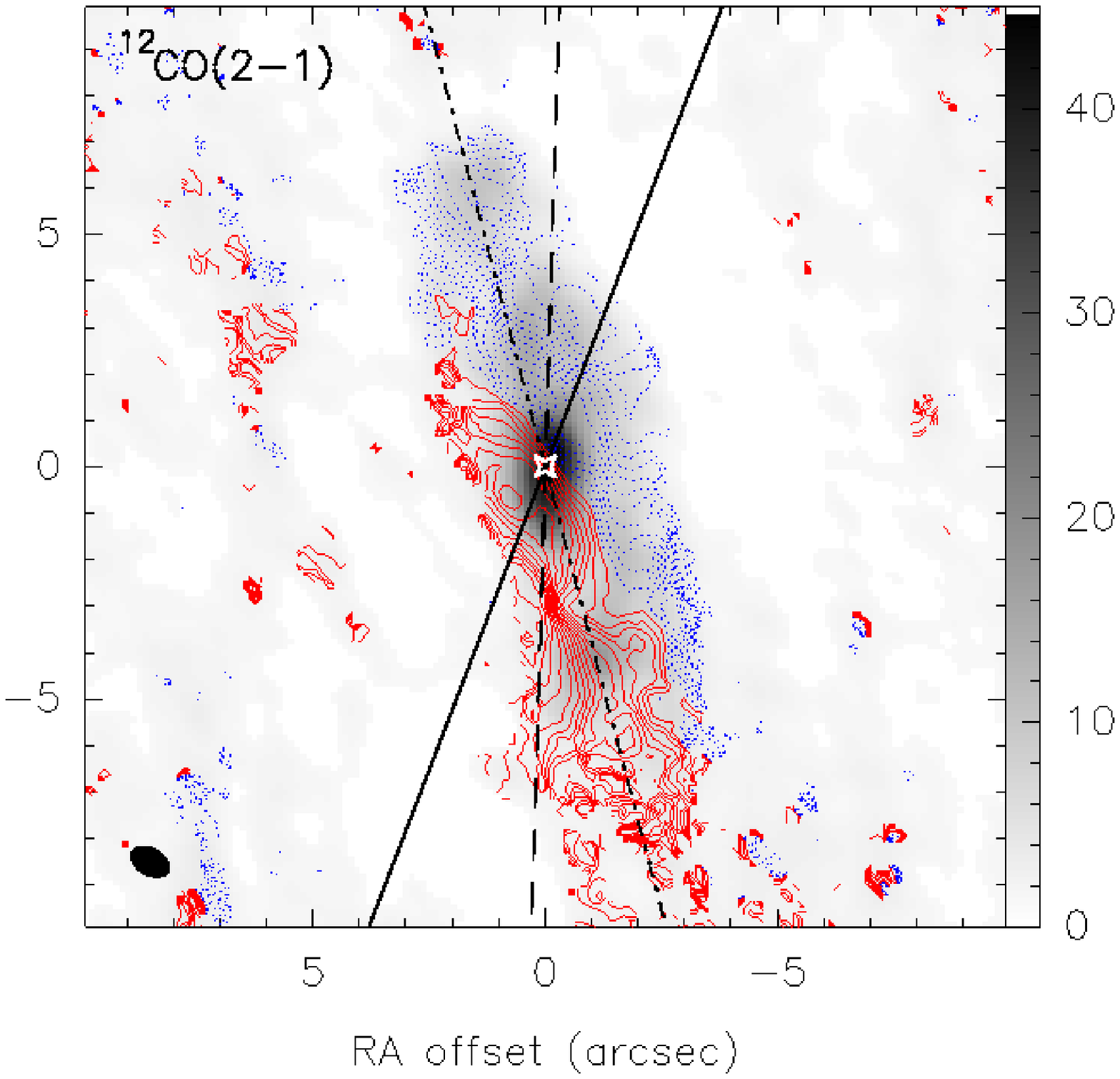}
}
\caption{
\textit{Left panel}: 
Overlay of the integrated $^{12}$CO(1--0) emission, same as 
Fig. \ref{fig:co10-21} (left panel), with CO mean-velocity field 
in contours spanning the range -180 to 180 km s$^{-1}$ 
in steps of 10 km s$^{-1}$. 
The white star indicates the dynamical center of the galaxy.
The velocities are referred to $V_{\rm sys, hel}$ = 744 km\,s$^{-1}$, 
solid (red) lines are used for positive velocities, and 
dashed (blue) lines for negative velocities.
The dashed line indicates the position angle of the major axis 
of the whole observed region (PA = 178$^{\circ}$ $\pm$ 1$^{\circ}$), while
the dot-dashed line traces the position angle of the major axis
of the bar-like structure (PA = 14$^{\circ}$ $\pm$ 2$^{\circ}$).
The continuum line indicates the position angle of the primary stellar bar 
identified with the NIR $H$-band 2MASS image (PA = $-21^{\circ}$) 
[see later the left panel of Fig. \ref{fig:h-band} and Sect. \ref{sec:nir}].
\textit{Right panel}: Same for $^{12}$CO(2--1). 
The dashed line indicates the position angle of the major axis 
of the whole $^{12}$CO(1--0) observed region (PA = 178$^{\circ}$ $\pm$ 1$^{\circ}$, see
left panel), while the dot-dashed line traces the position angle of the 
major axis of the $^{12}$CO(2--1) bar-like structure 
(PA = 15$^{\circ}$ $\pm$ 2$^{\circ}$).
The continuum line indicates the position angle of the primary stellar bar 
identified with the NIR $H$-band 2MASS image (PA = $-21^{\circ}$) 
[see later Fig. \ref{fig:h-band} and Sect. \ref{sec:nir}].
\label{fig:co-vel}
}
\end{figure*}

The $^{12}$CO distribution found here agrees well
with previous molecular gas maps, such as that given by \citet{regan01} and
\citet{helfer03} in the context of the BIMA SONG 
survey and that obtained by \citet{kuno07} with the 45\,m telescope
of the Nobeyama Radio Observatory. 
Although we only observed the inner $\sim$2\,kpc of the galaxy, 
the good PdBI resolution allows us to investigate the 
nuclear molecular gas distribution in \nnn\ more in detail than in 
BIMA SONG survey (typical resolution of $\sim$6\arcsec) and with 
the 45\,m NRAO telescope (FWHM$\sim$15\arcsec).   
The $^{12}$CO distribution is completely different from 
the ringed \hi\ morphology which exhibits an inner hole 
where instead the molecular gas is located \citep[][]{haan08,walter08}. 

Applying Eq. (\ref{h2mass}) to combined PdBI+30\,m 
data, we derived a total H$_{2}$ mass of 
M$\rm_{H_{2}}$$\sim$6.0$\times 10^{8}~\rm{M_{\odot}}$ 
($S\rm_{CO} = 668$\,Jy km\,s$^{-1}$, see Table \ref{table2}) 
within the 42\arcsec\ primary beam field of the PdBI.
Taking into account the mass of helium, the total molecular mass is 
M$\rm_{mol}$$\sim$8.2$\times10^{8} \rm{M_{\odot}}$.
This is roughly 63\% of the molecular gas mass within a 50\arcsec\
diameter (see Sect. \ref{sec:30m}).
The $\sim$18\arcsec\ $^{12}$CO(1--0) bar-like structure contributes 
an H$_{2}$ mass of M$\rm_{H_{2}}$$\sim$2.1$\times10^{8}~\rm{M_{\odot}}$, 
roughly one-third
of the H$_{2}$ mass computed within 42\arcsec, although
the feature occupies an area of only $\sim$5\% of the 42\arcsec\ beam.
\nnn, compared to other NUGA galaxies, is not particularly massive in molecular gas,
especially with respect
to the extraordinary case of NGC\,1961 with an H$_{2}$ mass of 
$\sim$1.8$\times10^{10}~\rm{M_{\odot}}$ \citep{francoise09}.

\begin{figure*}
\centering
\includegraphics[width=0.4\textwidth,angle=-90]{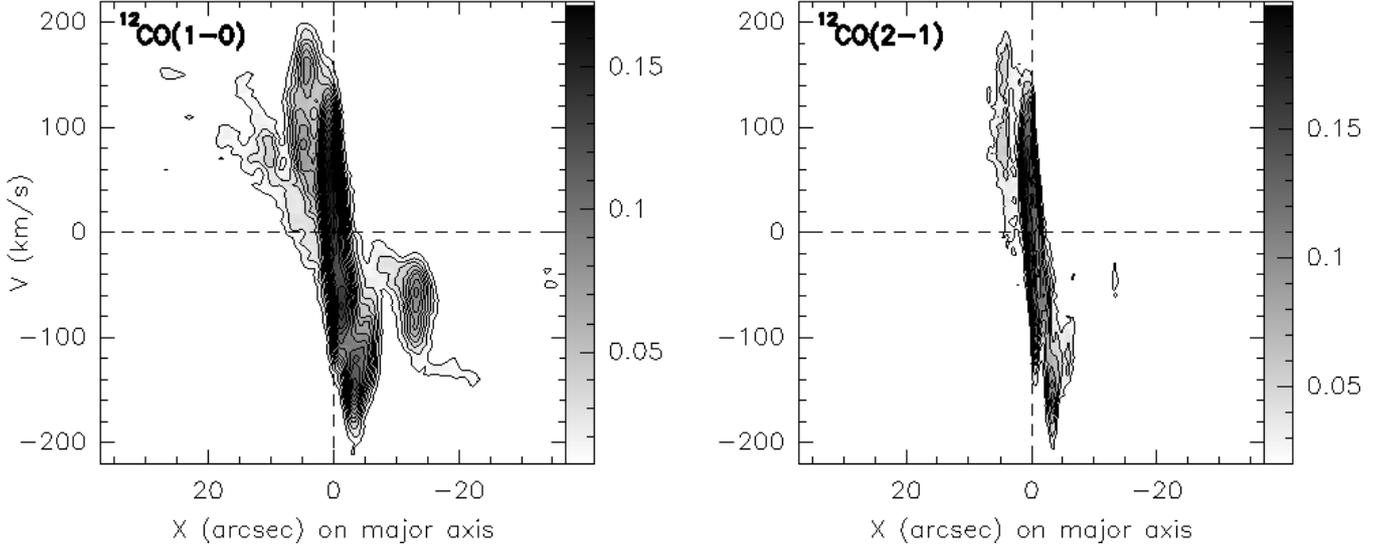}
\caption{\textit{Left panel}: $^{12}$CO(1--0) position-velocity
diagram along the major axis (PA = 178$^\circ$) of \nnn\
using the velocity range from -220 to 220 km s$^{-1}$ contoured 
over a grey-scale representation.
Contour levels are from 3$\sigma$ to 48$\sigma$ in steps of
3$\sigma$ ($\sigma$=3.6\,mJy\,beam$^{-1}$). 
The velocities are relative to $V_{\rm sys, hel}$ (= 744 km\,s$^{-1}$) and 
X are the offsets along the major axis in arcsecs.
\textit{Right panel}: The same for $^{12}$CO(2--1). 
Contour levels are from 3$\sigma$ to 30$\sigma$ 
in steps of 3$\sigma$ ($\sigma$=6.7\,mJy\,beam$^{-1}$).
}
\label{pv-major}
\end{figure*}

\begin{figure*}
\centering
\includegraphics[width=0.4\textwidth,angle=-90]{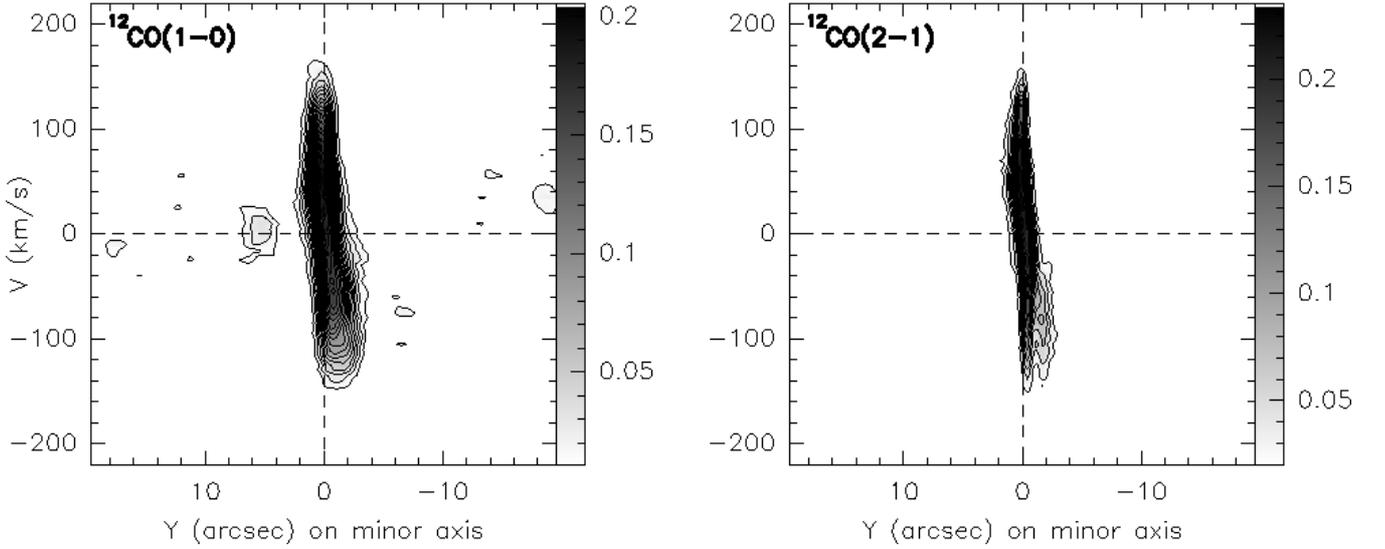}
\caption{\textit{Left panel}: Same as Fig. \ref{pv-major}
along the minor axis (PA = 88$^\circ$) of \nnn.
Contour levels are from 3$\sigma$ to 57$\sigma$ in steps of
3$\sigma$ ($\sigma$=3.6\,mJy\,beam$^{-1}$). 
Y are the offsets along the minor axis in arcsecs.
\textit{Right panel}: The same for $^{12}$CO(2--1). 
Contour levels are from 3$\sigma$ to 35$\sigma$ 
in steps of 3$\sigma$ ($\sigma$=6.7\,mJy\,beam$^{-1}$).
}
\label{pv-minor}
\end{figure*}

\begin{figure*}
\centering
\hbox{
\includegraphics[width=0.30\textwidth,angle=-90]{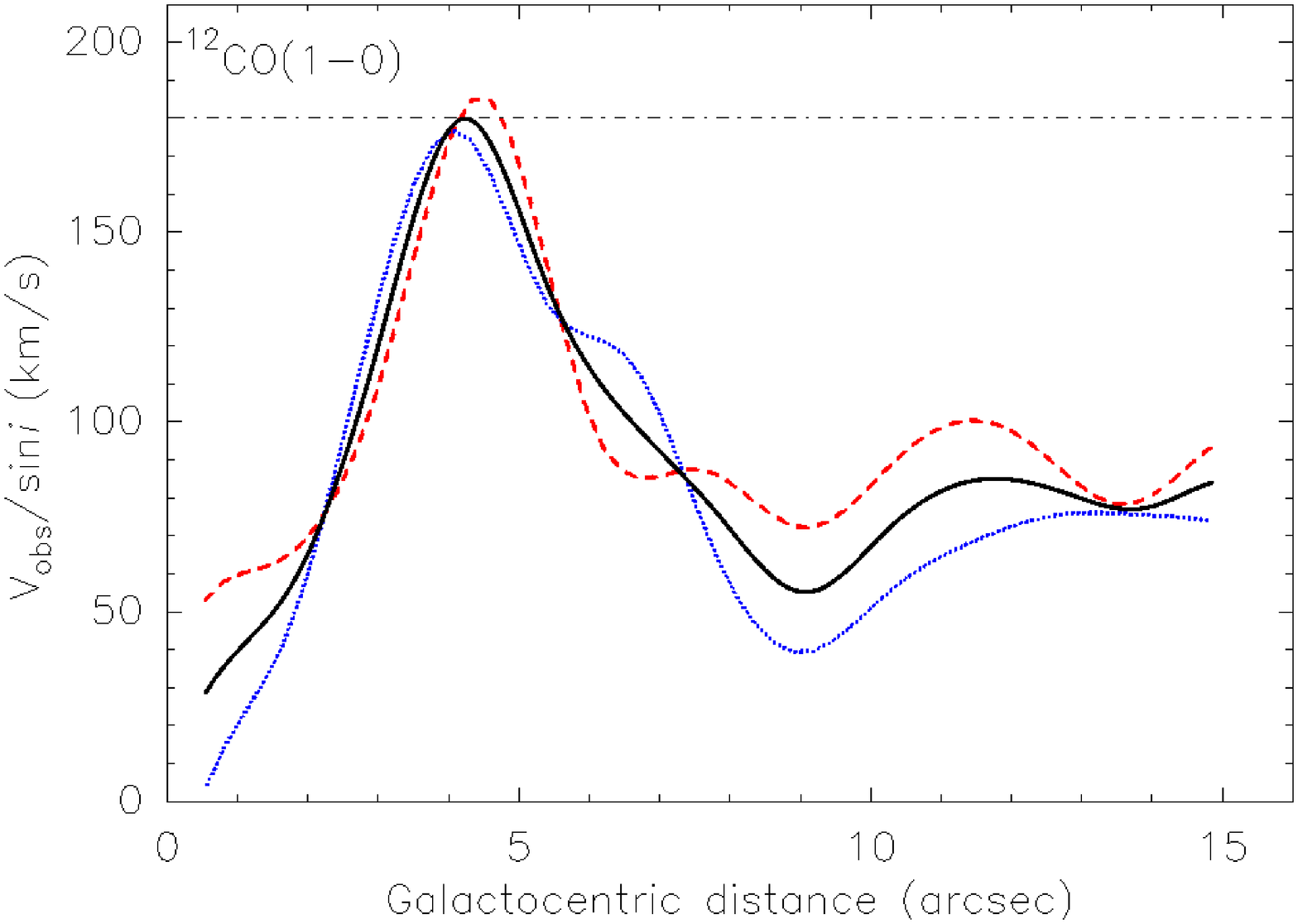}
\hspace{0.1\textwidth}
\includegraphics[width=0.30\textwidth,angle=-90]{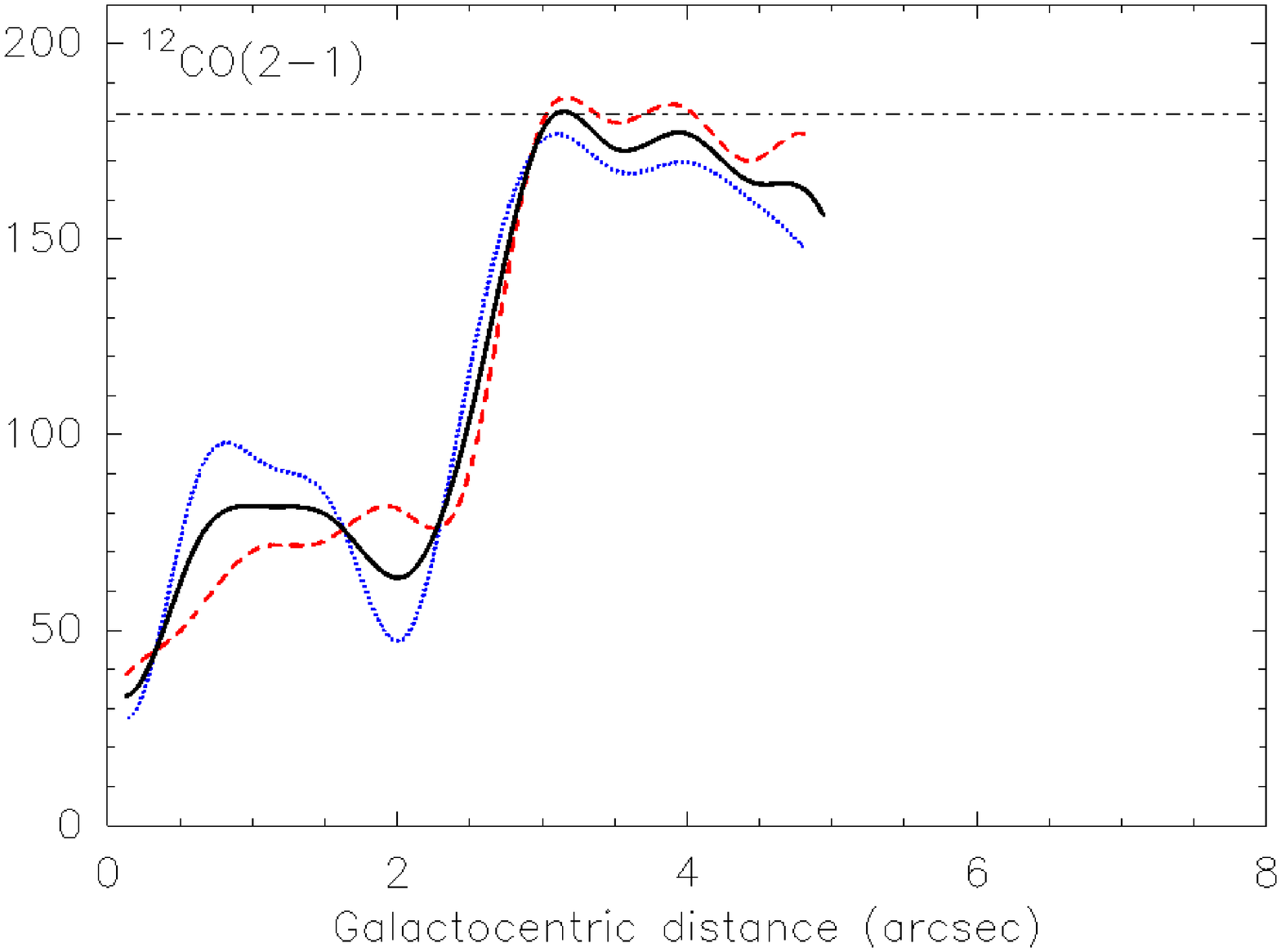}
}
\caption{\textit{Left panel}:
The $^{12}$CO(1--0) RC derived as described in the text.
The positive (negative) velocities, corrected for inclination, are shown as 
a red dashed (blue dotted) line; the black line shows the average. The horizontal dot-dashed
line at 180 km s$^{-1}$ indicates the velocity of the bulk of the molecular 
gas.
\textit{Right panel}:
The same as the left panel, but for $^{12}$CO(2--1).
}
\label{fig:rc}
\end{figure*}

\subsection{CO(2--1)/CO(1--0) line ratio \label{sec:coratios}}
Information about the local excitation conditions of the molecular gas
can be inferred from the line ratio R$_{\rm 21}$=I$_{\rm 21}$/I$_{\rm10}$.
This ratio is obtained by comparing the $^{12}$CO maps of the two transitions,
at the same resolution and with the same spatial frequency sampling.
Figure \ref{fig:ratio} shows R$_{\rm 21}$ ratio with $^{12}$CO(1--0) contours 
as in Fig. \ref{fig:co10-21} (left panel).
In the observed region, the line ratio ranges from 0.25 to 1 but 
the bulk of the emission has a ratio between 0.4 and 0.7.
These R$_{21}$ line ratio values are consistent with R$_{21}$ = 0.6
obtained by \citet{melanie08}, and more in general with optically thick emission 
in spiral disks \citep[e.g.,][]{braine92,santi93}.
The R$_{21}$ peaks of $\sim$1 are reached in the center of \nnn\
and at the southern extreme of the elongated $^{12}$CO emission region.
A higher  excitation of the molecular gas in the nucleus, suggested
by a higher R$_{21}$ line ratio, is consistent with the 
HCN(1--0) emission in the same region (see Sect. \ref{sec:30m}).

\subsection{CO Kinematics \label{sec:cokinematics}}
Figures \ref{channels10}  and \ref{channels21} show the 
velocity-channel maps of $^{12}$CO(1--0) and $^{12}$CO(2--1) 
emission, respectively, in the central region of \nnn. 
The inner $^{12}$CO emission of the galaxy exhibits 
signatures of non-circular motions both at negative and positive velocities. 
These non-circular components are associated both with the 18\arcsec\ bar-like 
structure and the spiral feature detected beyond the bar-like structure and will 
be discussed in detail later, in Sect. \ref{sec:corc}, where we analyze the rotation 
curve derived with our $^{12}$CO data.

$^{12}$CO(1--0) isovelocity contours (first-moment map) are superposed 
on the $^{12}$CO(1--0) integrated intensity in Figure \ref{fig:co-vel}
(left panel).
The white star indicates the dynamical center of the galaxy, assumed 
coincident with the phase tracking center of our observations, and the 
velocities are relative to the systemic heliocentric velocity, 
$V_{\rm sys, hel}$ = 744 km\,s$^{-1}$ (see Sect. \ref{sec:dyncen}).
The dashed line traces the position angle of the major axis 
of the observed region, PA = 178$^{\circ}$ $\pm$ 1$^{\circ}$
(almost vertical), obtained by maximizing the symmetry in the position 
velocity diagrams.
This position angle is close to that of the entire galaxy, 
as given by the surface brightness profiles 
(172$^\circ$, see Sect. \ref{sec:nir}),
and given in the Uppsala General Catalog (173$^{\circ}$). 
The dot-dashed line traces the position angle of the major axis
of the bar-like structure, PA = 14$^{\circ}$ $\pm$ 2$^{\circ}$.
The right panel of Fig. \ref{fig:co-vel} shows $^{12}$CO(2--1) 
isovelocity contours on the $^{12}$CO(2--1) integrated intensity, 
where the position angle of the major axis of the observed region 
is very similar to that found for the $^{12}$CO(1--0) inner region 
(PA = 15$^{\circ}$ $\pm$ 2$^{\circ}$).
The continuous lines, in the left and right panels of 
Fig. \ref{fig:co-vel}, show the PA of the stellar bar 
identified in the NIR with the $H$-band 2MASS image (PA = $-21^{\circ}$) 
[see below the left panel of Fig. \ref{fig:h-band} and Sect. \ref{sec:nir}].
The different PAs of the stellar bar and the molecular gas bar-like structure 
suggest that the gas is leading the stellar bar.
Signatures of non-circular motions are visible also in Fig. \ref{fig:co-vel},
since the isovelocity contours appear severely tilted along the 
molecular gas bar-like structure both in $^{12}$CO(1--0) and 
$^{12}$CO(2--1).

Figures \ref{pv-major} and \ref{pv-minor} show 
position-velocity (p-v) cuts along the major (PA = 178$^\circ$) and 
minor axis (PA = 88$^\circ$) of \nnn, respectively.
In both figures, the $^{12}$CO(1--0) emission is given in the left panel 
and $^{12}$CO(2--1) in the right.
Our p-v plots along major axis are consistent with the $^{12}$CO(1--0) 
p-v diagram obtained by \citet{regan02}, where the different velocity 
range between nuclear region and bar ends is yet clearer, because 
they mapped in $^{12}$CO(1--0) the whole galaxy.
\citet{regan02}'s p-v diagram (Fig. 5a in their paper) shows that 
the components we found at 10--15\arcsec\ from the center 
at velocities of 60--80\,km\,s$^{-1}$ (not corrected for inclination) 
maintain this velocity also at larger radii (until $\sim$50\arcsec\ 
from the center), and then again increase and reach the velocity 
assumed by bar ends.

\subsection{CO rotation curve and dynamical mass \label{sec:corc}}
We have derived a rotation curve (RC) from
the p-v diagram along the kinematic major axis of \nnn\ 
(PA = 178\,$^\circ$). 
By fitting multiple gaussian profiles to the spectra across the major 
axis we calculated the terminal velocities, and the fitted velocity 
centroids, corrected for inclination ($\sin i$, $i=$61\fdg3), 
give $V_{\rm obs}/\sin i$ for each galactocentric distance. 
Figure \ref{fig:rc} shows RCs for $^{12}$CO(1--0) [left]  
$^{12}$CO(2--1) [right], where for each line we plotted both 
the two curves derived from either side of the major axis and 
their combination into a average by spline interpolation.
This interpolation is justified by the similar behavior of the  
positive velocity curve and the negative velocities for both lines.

The $^{12}$CO(1--0) RC reaches a maximum of 180\,km\,s $^{-1}$ 
(velocity corrected for inclination)
at r$\sim$4\farcs3 ($\sim$0.2\,kpc), and then decreases until velocities between 
$\sim$60\,km\,s $^{-1}$ and $\sim$85\,km\,s$^{-1}$ maintaining these velocities
until $\sim$15\arcsec\ ($\sim$0.74\,kpc) from the center (Fig. \ref{fig:rc}, left panel).
The $^{12}$CO(2--1) RC shows a similar maximum velocity, $\sim$183\,km\,s$^{-1}$, reached 
at r$<$5\arcsec (Fig. \ref{fig:rc}, right panel).   
At larger distances from the nucleus (r$>$15\arcsec), 
we expect that the RC again 
increases until $\sim$180\,km\,s$^{-1}$, the velocity at the ends of the bar, 
consistently with p-v plots mapping the whole galaxy \citep[e.g.,][]{regan02}.
Asymmetries are also seen in the H$\alpha$ RC of \nnn\ \citep{chemin03}, 
both in the inner disk (between 25\arcsec and 34\arcsec from the nucleus, 
already outside the range of our $^{12}$CO observations) and at larger 
radii ($>$88\arcsec). 
The form of the RC of \nnn, with remarkable dips in velocity near the nucleus, 
is not unusual for galaxies with circumnuclear bars or gas disks \citep[e.g.,][]{rubin97}.

Such behavior often reflects non-circular motions depending both on the 
internal structure of the galaxy (e.g., the bar and consequent streaming motions) 
and the orbital parameters of the interaction with nearby companions.
Another important source of non-circular motions may be the kinematic 
feedback to the gas from star formation \citep{beauvais99}.
Regions of intense SF and turbulent motions 
in those regions may lead to an increase in the local velocity dispersion.
In \nnn, the dips may be due to all three mechanisms, a
combination of the effects of the interaction, the strong bar, and
the kinematic feedback to the gas from SF. 

\begin{figure*}
\centering
\includegraphics[width=0.43\textwidth,angle=-90]{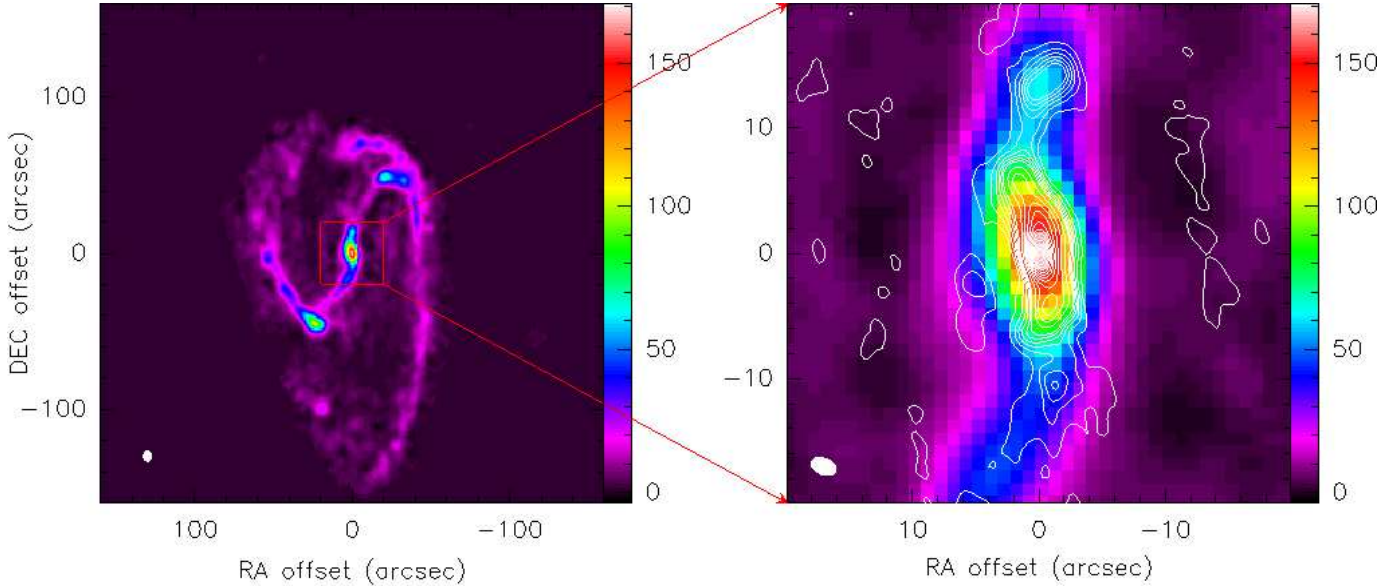}
\caption{\textit{Left panel}: The BIMA SONG $^{12}$CO(1--0) emission image of \nnn\
\citep{regan01}.
The beam of 6\farcs6 $\times$ 5\farcs5 is plotted in the lower left.
The inner 320\arcsec are shown. 
\textit{Right panel}: 
NUGA $^{12}$CO(1--0) contours (in white) as in Fig. \ref{fig:co10-21}  
(left panel) overlaid on the BIMA SONG $^{12}$CO(1--0) emission image.
The $^{12}$CO(1--0) NUGA beam of 2\farcs1 $\times$ 1\farcs3 
is plotted in the lower left.
The inner 40\arcsec are shown.
}
\label{fig:bima}
\end{figure*}  

\begin{figure*}
\centering
\hbox{
\includegraphics[width=0.5\textwidth,angle=0]{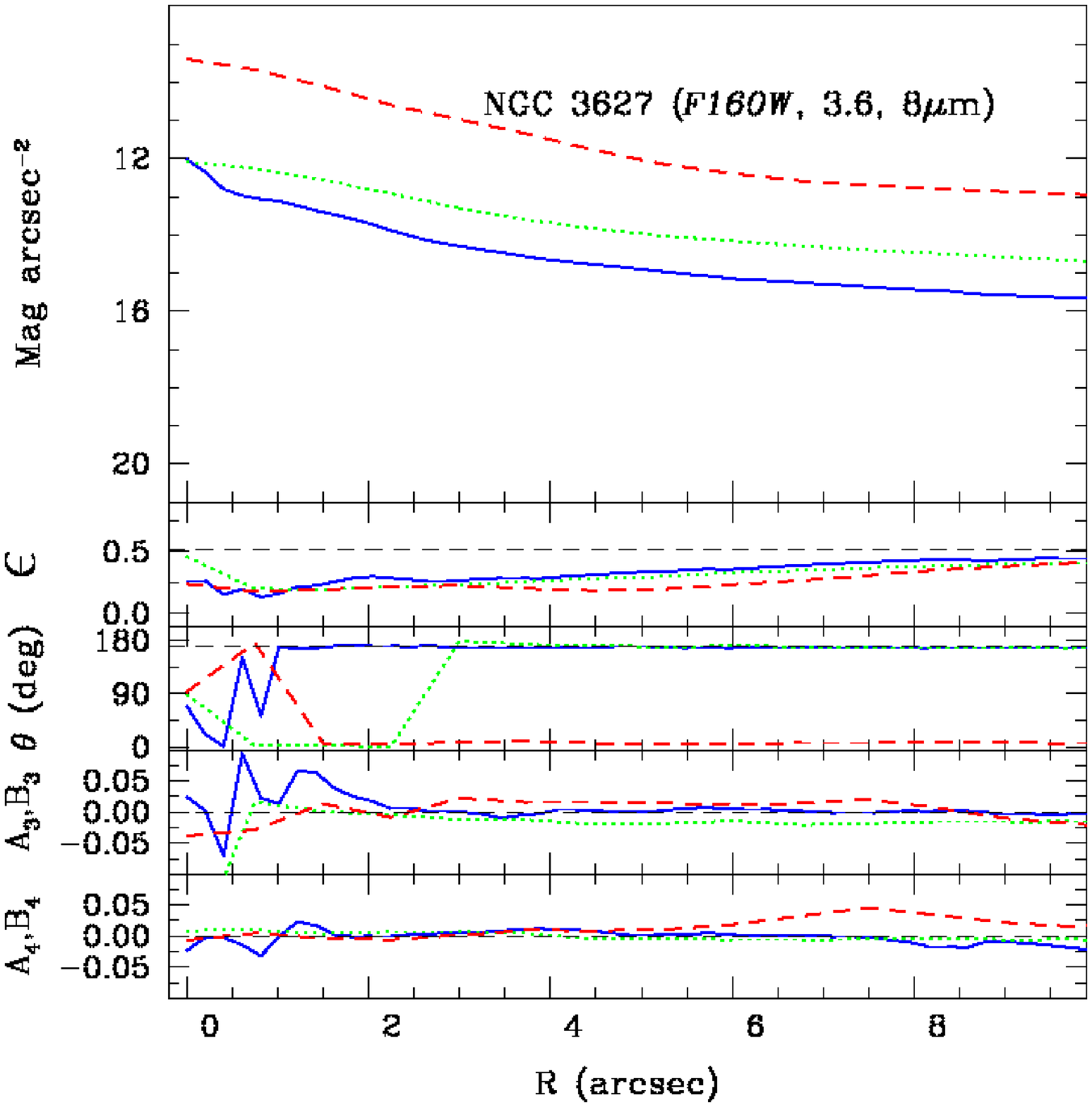}
\includegraphics[width=0.5\textwidth,angle=0]{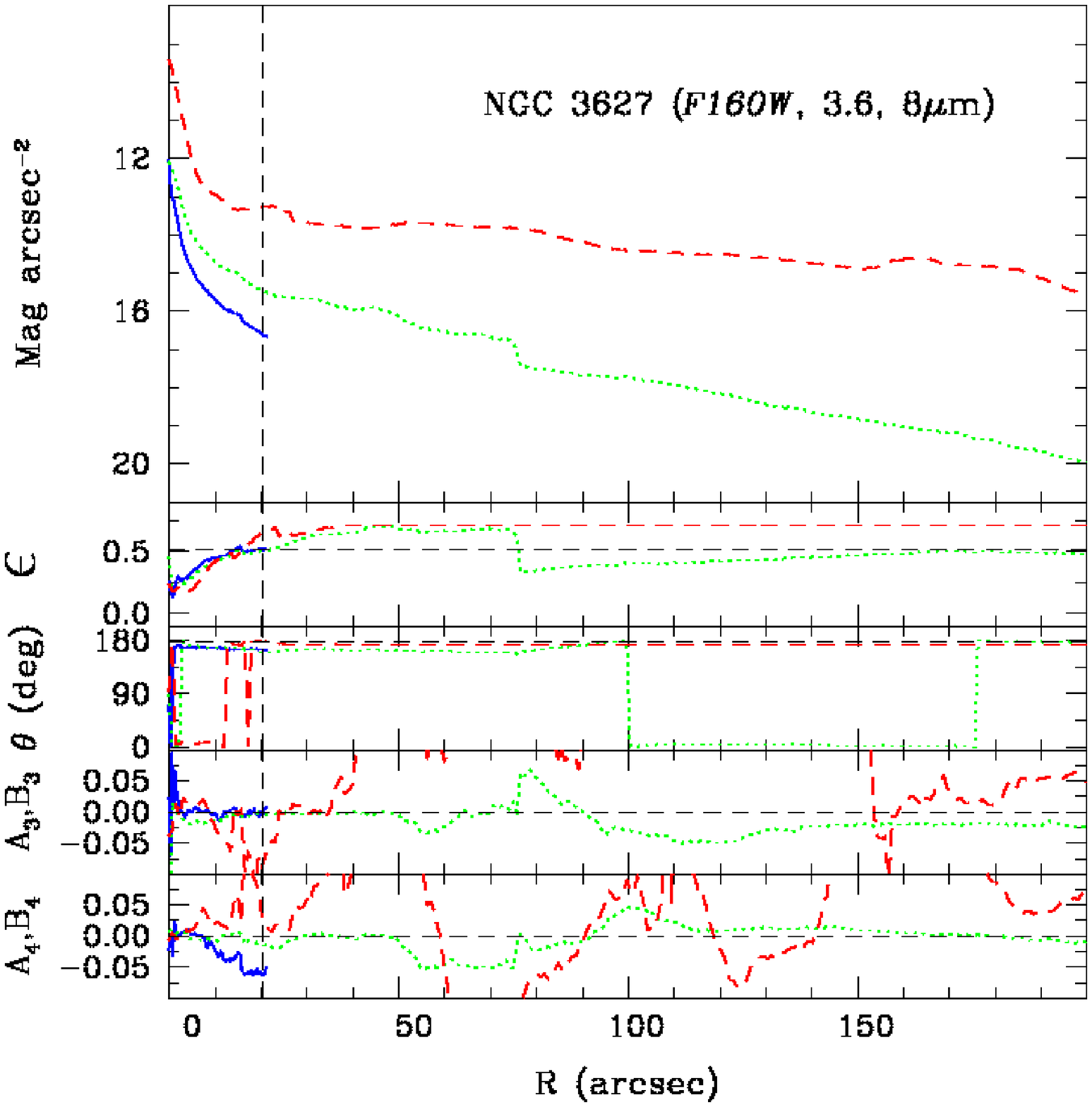}
}
\caption{The upper panels show radial surface brightness profiles of \nnn\ 
made by fitting elliptical isophotes.
The left panel shows a smaller FOV than the right panel.
The solid (blue) line corresponds to \hst\ F160W, the dotted (green) line
to IRAC 3.6\,\micron, and the dashed (red) line to the IRAC 8\,\micron.
The lower panels display the radial runs of ellipticity $\epsilon$, 
position angle $\theta$, and $\cos(4\theta)$ terms in the ellipse fitting 
residuals.
The adopted PA (178$^\circ$, see Sect. \ref{sec:cokinematics}) and the 
best-fit inclination (61.3$^\circ$) are shown by dashed horizontal 
lines in the lower panels.
The dashed vertical line shows a radius of 1\,kpc (20\farcs4).
\label{fig:profiles}
}
\end{figure*}

From the RC, we can estimate the dynamical mass within a certain radius 
with the formula 
$M(R) = 2.325\times 10^5\ \alpha\ R V^2(R)$
where $M(R)$ is in $\rm{M_{\odot}}$, $R$ in kpc, $V$ in km\,s$^{-1}$,
and $\alpha$ is a factor related to the geometry.
The choice of $V$ to use in the above formula is complicated 
by the presence of dips in the RC (Fig. \ref{fig:rc}). 
Although these dips are due to non-circular motions, 
at larger radii both CO \citep[e.g.,][]{regan02} and \hi\ 
\citep[e.g.,][]{zhang93,haan08} RCs are flat and at the same 
velocity found at the bar ends. 
We can thus estimate the dynamical mass 
using the maximum velocity of 180\,km\,s$^{-1}$ (corrected for inclination) 
reached at $\sim$210\,pc from the center.

Assuming for $\alpha$ a value of 0.8, intermediate between the value 
appropriate for a spherical distribution (1.0) and that for a
flat disk (0.6), the above formula gives a dynamical mass of
M$_{\rm dyn}=1.3\times10^{9}$\,M$_\odot$ 
within a radius of 4\farcs3 ($\sim$0.2\,kpc).
Continuing to neglect non-circular motions and assuming a roughly flat 
RC also at larger radii, the dynamical mass should be 
M$_{\rm dyn}=6.0\times10^{9}$\,M$_\odot$
within a radius of 
21\arcsec\ ($\sim$1\,kpc).
In the same region, we estimate a H$_{2}$ mass of 
$\sim$6.0$\times10^{8}$\,M$_\odot$ (see Sect. \ref{sec:codistribution}), 
$\sim$10\% of the dynamical mass.
\citet{zhang93} found a dynamical mass of 
M$_{\rm dyn}=5.8\times10^{9}$\,$M_\odot$ (value scaled to our adopted 
distance of $D=10.2$\,Mpc) within a radius of 23\arcsec, a value consistent with
our determination in a similar region, considering uncertainties from a
different assumed inclination and rotation velocity. 
Moreover, \citet{zhang93} found a ratio 
M$_{\rm H_{2}}$/M$_{\rm dyn}$$\sim$11\% within the radius of 
23\arcsec, similar to our value. 
Nevertheless, because of the clear signature of non-circular motions 
in the RC, our estimate of dynamical mass is very uncertain, probably $\pm$50\%.

\section{Comparison with other data\label{sec:comparison}}
Here we present a comparison of our  
$^{12}$CO observations and images at other wavelengths available 
for \nnn.
These comparisons allow both to assess possible correlations between 
different tracers of the ISM and to determine the location of the 
dynamical resonances, useful for probing gas inflow in 
the circumnuclear region of the galaxy.
All images have been centered on the phase tracking center of 
our $^{12}$CO interferometric observations (see Table \ref{table1}), 
and -when necessary- properly rotated with North up and E left.

\subsection{Another CO map: NUGA vs. BIMA SONG\label{sec:co_comparison}}
The left panel of Figure \ref{fig:bima} shows the BIMA SONG $^{12}$CO(1--0) 
emission image of \nnn. 
This image reveals, like in the optical, a nuclear barred structure and a 
pronounced and asymmetric spiral pattern. 
A close-up of the inner 40\arcsec of the  BIMA SONG $^{12}$CO(1--0) map 
with our $^{12}$CO(1--0) contours overlaid, as in 
Fig. \ref{fig:co10-21} (left panel), is shown in the right panel of 
Fig. \ref{fig:bima}.
The two $^{12}$CO data sets agree quite well;
the only significant disagreement is present toward the south, where 
our higher-resolution $^{12}$CO(1--0) contours delineate a barred structure that develops 
mainly toward the south, while the BIMA SONG $^{12}$CO(1--0) emission  
is more rotated toward the east/south-east.

\subsection{Near- and mid-infrared emission \label{sec:nir}}
NIR images of \nnn\ have been used both to derive the surface brightness 
profiles and to perform comparisons with our $^{12}$CO observations.
We compared the surface brightness profiles of the ground-
and space-based images by extracting elliptically averaged profiles,
centered on the brightness peaks.
The position angle and ellipticity were allowed to vary in the ellipse fitting. 
These radial profiles are shown in Figure \ref{fig:profiles}, where 
the dashed horizontal lines in the lower panels correspond to the
adopted position angle (PA) and inclination (\textit{i}).
The three profiles (\hst\ F160W represented by the solid (blue) line, 
IRAC 3.6\,\micron\ by the dotted (green) line, and IRAC 8\,\micron\ 
by the dashed (red) line) are quite similar, in particular 
the \hst\ F160W and the IRAC 3.6\,\micron\ profiles show a similar 
trend within the inner 10\arcsec.
Also shown are the runs of ellipticity $\epsilon$, ellipse position 
angle $\theta$, and $\cos(4\theta)$ residuals of the ellipse fitting.
The adopted PA is 178$^\circ$ (see Sect. \ref{sec:cokinematics}),
similar to the fitted PA of 172$^\circ$.
The inclination computed from the elliptical fits of the
IRAC 3.6\,\micron\ image converges 
to 61.3$^\circ$, very close to the inclination given by NED, 62.5$^\circ$.

\begin{figure}
\centering
\includegraphics[width=0.45\textwidth,angle=-90]{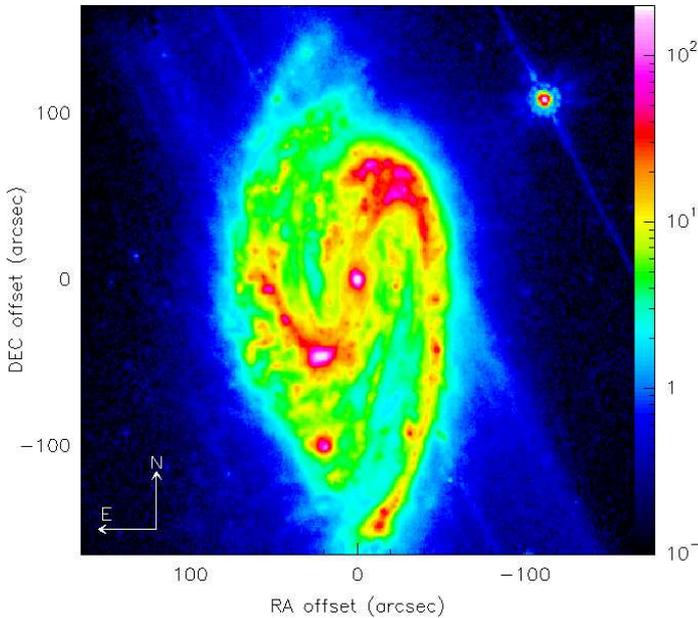}
\caption{
The large-scale IRAC 8\,$\mu$m image of \nnn\, 
centered on the phase tracking center of our $^{12}$CO interferometric 
observations.
The inner $330^{\prime\prime}$ are shown.
}
\label{fig:8mu}
\end{figure}

\begin{figure*}
\centering
\includegraphics[width=0.43\textwidth,angle=-90]{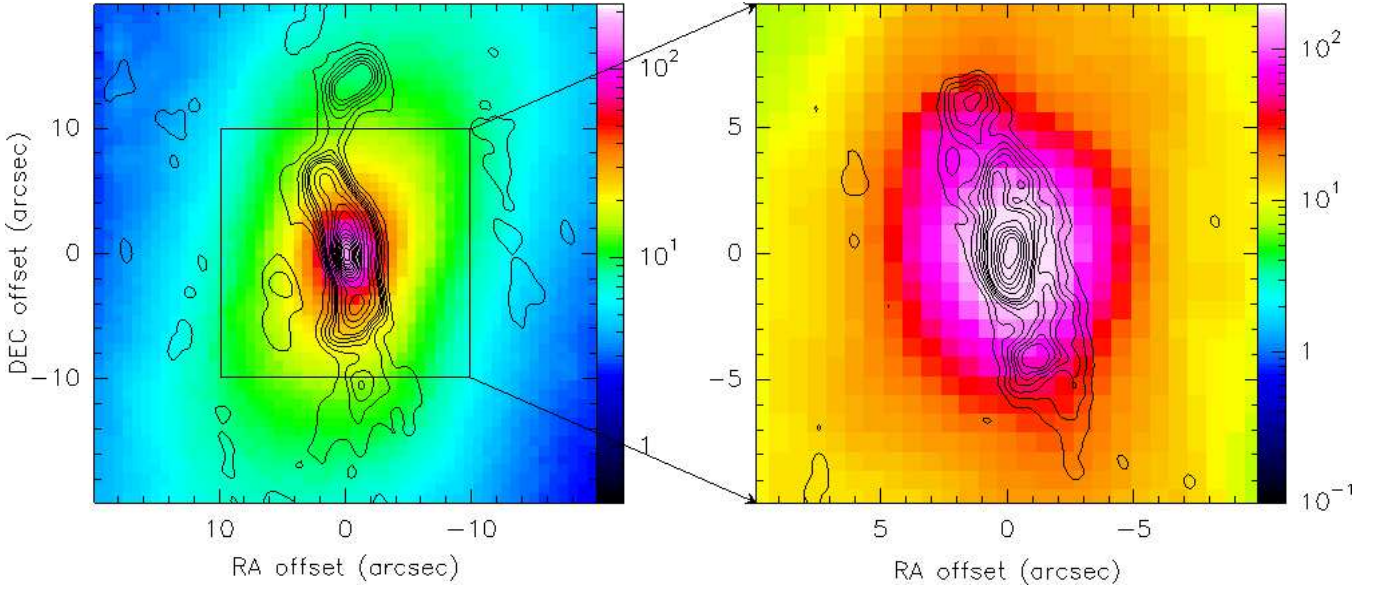}
\caption{\textit{Left panel}:  
$^{12}$CO(1--0) integrated intensity contours as in Fig. \ref{fig:co10-21} 
(\textit{left panel}) overlaid on the IRAC 3.6\,$\mu$m image of 
\nnn\ in false color.
The inner 40\arcsec\ are shown. 
\textit{Right panel}: 
Same for $^{12}$CO(2--1) integrated intensity contours as in 
Fig. \ref{fig:co10-21} (\textit{right panel}). 
The inner 20\arcsec\ are shown. 
}
\label{fig:irac-3.6-co}
\end{figure*}

\begin{figure*}
\centerline{
\hbox{
\includegraphics[width=0.4\textwidth,angle=-90]{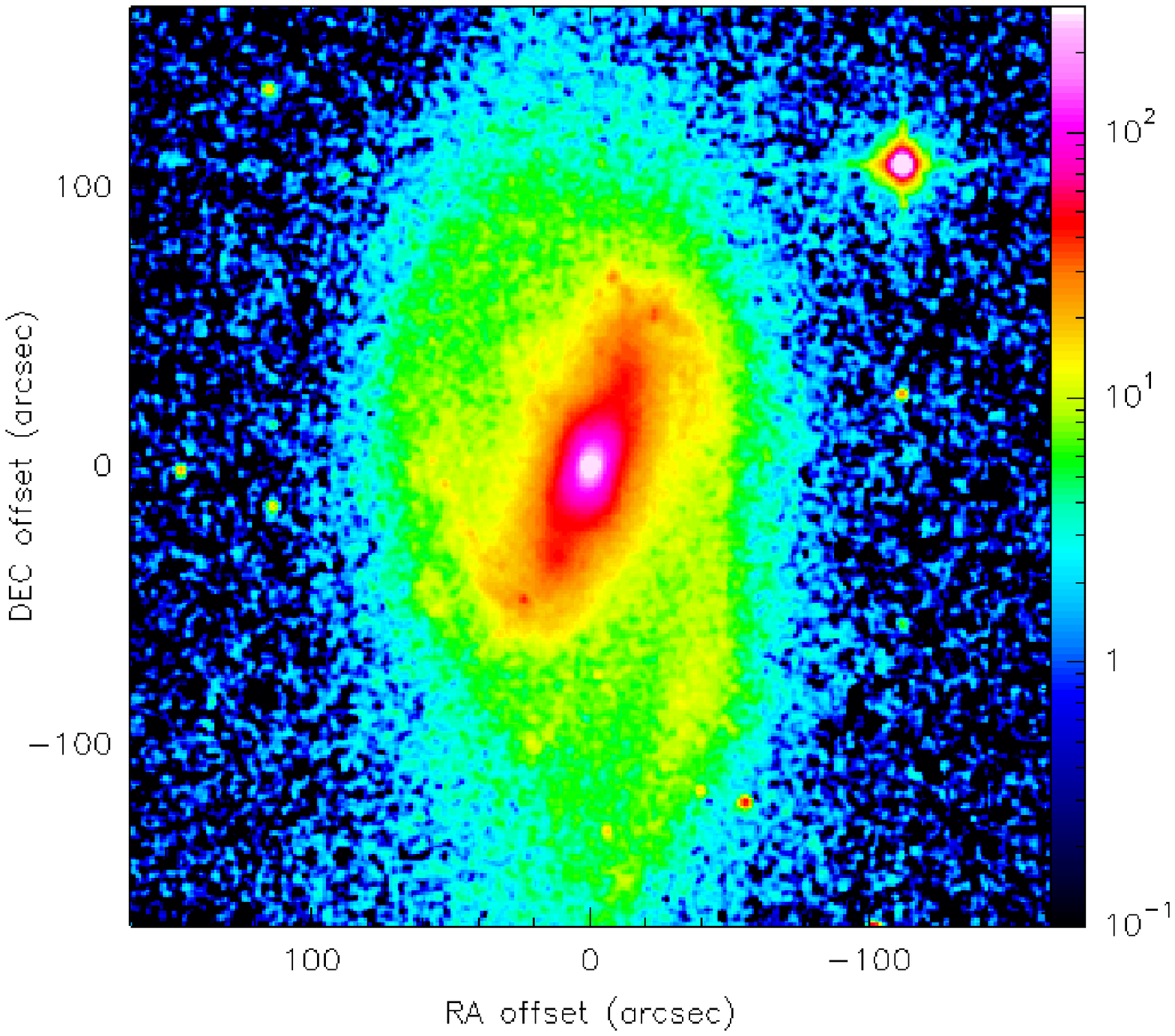}
\hspace{0.1\textwidth}
\includegraphics[width=0.4\textwidth,angle=-90]{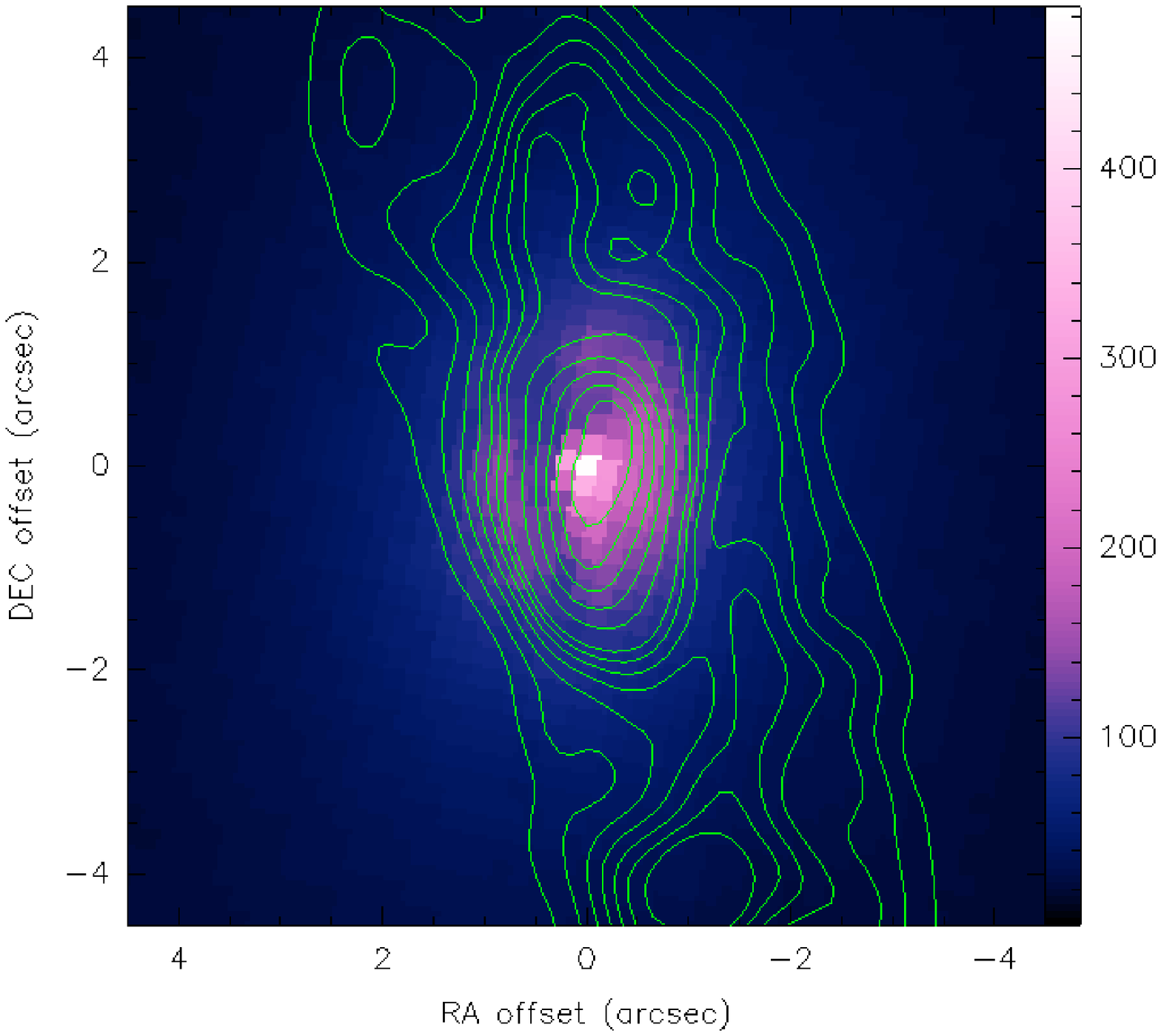}
}
}
\caption{
\textit{Left panel}: Large-scale $H$-band 2MASS image of \nnn\, centered on the 
phase tracking center of our $^{12}$CO interferometric observations.
The inner $330^{\prime\prime}$ are shown.
\textit{Right panel}: 
Inner 9\arcsec\ of the F160W/\hst\ $H$-band image of \nnn\, rotated with N up 
and E left and centered on the phase tracking center of our $^{12}$CO 
interferometric observations with overlaid NUGA $^{12}$CO(2--1) contours (in green) 
as in Fig. \ref{fig:co10-21} (right panel).
}
\label{fig:h-band}
\end{figure*} 

Figure \ref{fig:8mu} shows a large-scale view of \nnn\ at 8\,$\mu$m 
(\spitzer-IRAC).
The longest-wavelength 8\,$\mu$m IRAC band is dominated by 
PAHs and perhaps some hot dust continuum emission. 
This figure clearly reveals a spiral pattern, asymmetric with respect
to the major axis, and heavy dust lanes, signature of a strong 
density wave action. 
This image also shows an evident perturbed morphology of the eastern 
arm.
The bulk of the inner warm dust emission appears 
configured in a smooth elongated (north/south) disk,
but the large-scale bar seen in the BIMA-SONG CO image
is clearly reflected in the 8\,$\mu$m emission.

Figure \ref{fig:irac-3.6-co} shows the stellar morphology traced by 
the 3.6\,$\mu$m (\spitzer-IRAC) emission in the circumnuclear region 
of \nnn, with overlaid $^{12}$CO(1--0) [left panel] and 
$^{12}$CO(2--1) [right panel] intensity contours.
This comparison shows that the large-scale 3.6\,$\mu$m stellar bar 
and the  molecular gas bar-like feature have different orientation 
in the plane of the galaxy. 
The stellar bar has a PA of $-21^{\circ}$, 
while the molecular bar has a PA of 
$\sim$14$^{\circ}$ in $^{12}$CO(1--0) and $\sim$15$^{\circ}$ in $^{12}$CO(2--1) 
[see Sect. \ref{sec:cokinematics}].
This difference in orientation suggests that the molecular gas is leading 
the stellar bar.
The right panel of Fig. \ref{fig:profiles} shows clear signatures of a 
bar-like feature at 3.6\,$\mu$m at a radius of 60-70\arcsec: the surface 
brightness undergoes an inflection; the ellipticity dips; and the PA 
changes slightly. 
The bar would thus have a radius between 3 and 3.4\,kpc.

\begin{figure*}
\centering
\includegraphics[width=0.43\textwidth,angle=-90]{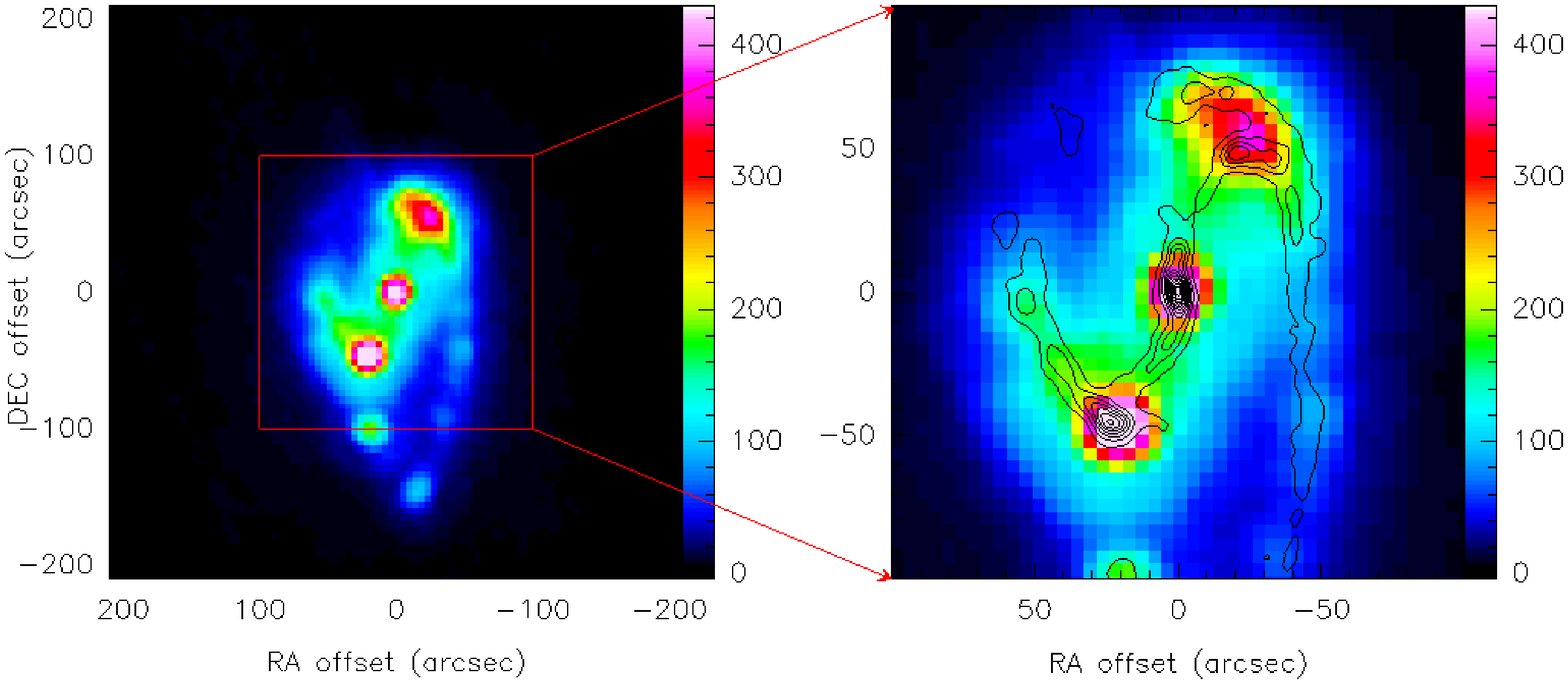}
\caption{\textit{Left panel}:  
Large-scale \spitzer-MIPS 70\,$\mu$m image of \nnn, centered on the phase 
tracking center of our $^{12}$CO interferometric observations. 
The inner $420^{\prime\prime}$ are shown.
The (red) box shows the central $200^{\prime\prime}$, displayed in the 
\textit{right panel}. 
\textit{Right panel}: 
BIMA $^{12}$CO(1--0) integrated intensity contours overlaid on the 
\spitzer-MIPS 70\,$\mu$m image of \nnn. 
The inner $200^{\prime\prime}$ are shown.
}
\label{fig:mips70-bima}
\end{figure*}  

\begin{figure*}
\centering
\includegraphics[width=0.43\textwidth,angle=-90]{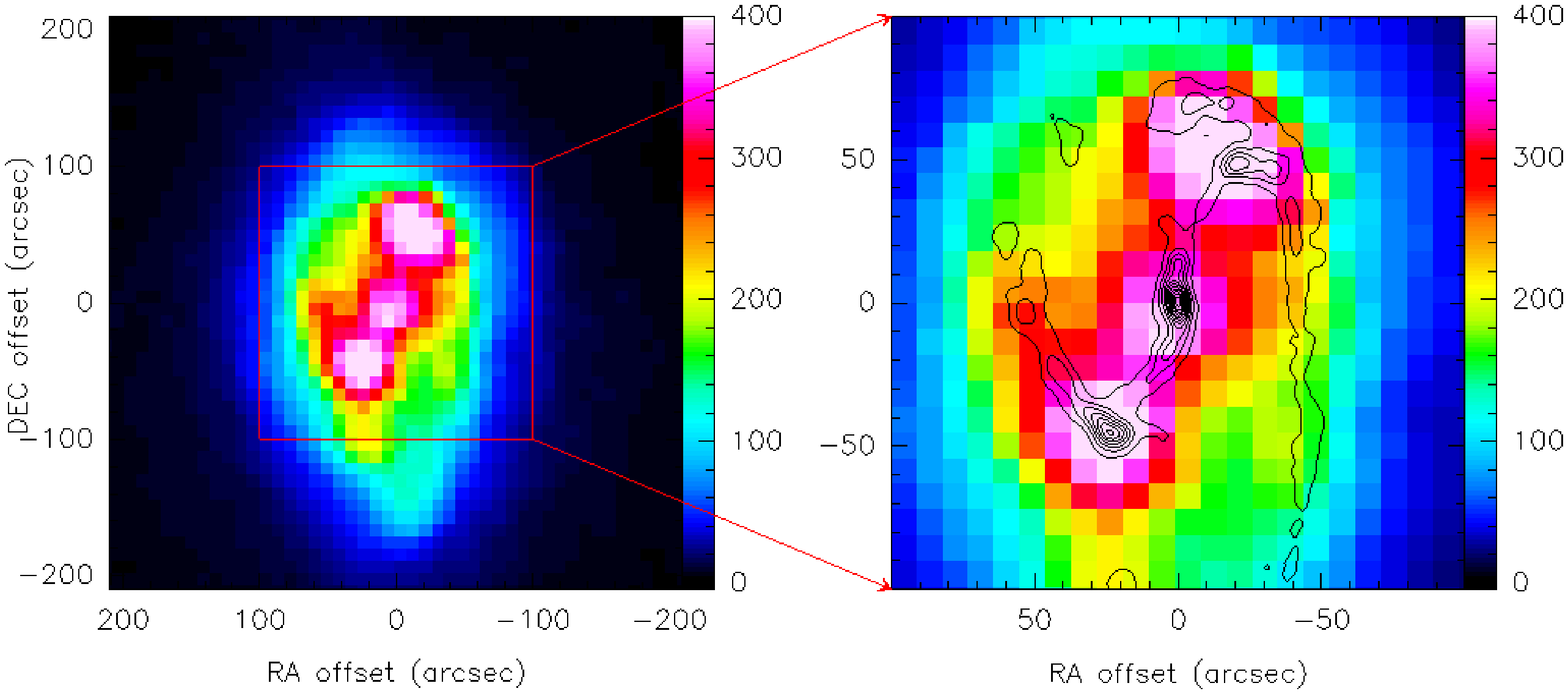}
\caption{\textit{Left panel}:  
Large-scale \spitzer-MIPS 160\,$\mu$m image of \nnn, centered on the phase 
tracking center of our $^{12}$CO interferometric observations. 
The inner $420^{\prime\prime}$ are shown.
The (red) box shows the central $200^{\prime\prime}$, displayed in the 
\textit{right panel}. 
\textit{Right panel}: 
BIMA $^{12}$CO(1--0) integrated intensity contours overlaid on the 
\spitzer-MIPS 160\,$\mu$m image of \nnn. 
The inner $200^{\prime\prime}$ are shown.
}
\label{fig:mips160-bima}
\end{figure*}  

The 1.6\,$\mu$m ($H$-band, 2MASS) large-scale morphology of \nnn\ is shown in 
the left panel of Figure \ref{fig:h-band}. 
Most of the stellar mass in a typical galaxy is locked up in cool stars, 
whose light is emitted longward of 1\,$\mu$m. 
Since the stellar spectrum tends to peak around 1.6\,$\mu$m, corresponding to the 
$H$-band NIR window, 1.6\,$\mu$m emission is an effective tracer of stellar mass. 
Like the emission at 3.6\,$\mu$m,
the 1.6\,$\mu$m morphology of \nnn\ is much smoother than that found 
at 8\,$\mu$m (Fig. \ref{fig:8mu});
the former maps trace the older stellar population,
while the latter is tracing the sites of star formation which tend to be clumpy,
with a more inhomogeneous distribution.
Like the 3.6\,$\mu$m emission, the $H$-band image shows a large-scale stellar bar
with PA = $-21^{\circ}$ and a radius of roughly 3-3.4\,kpc (60-70\arcsec).

\begin{figure*}
\centering
\includegraphics[width=0.43\textwidth,angle=-90]{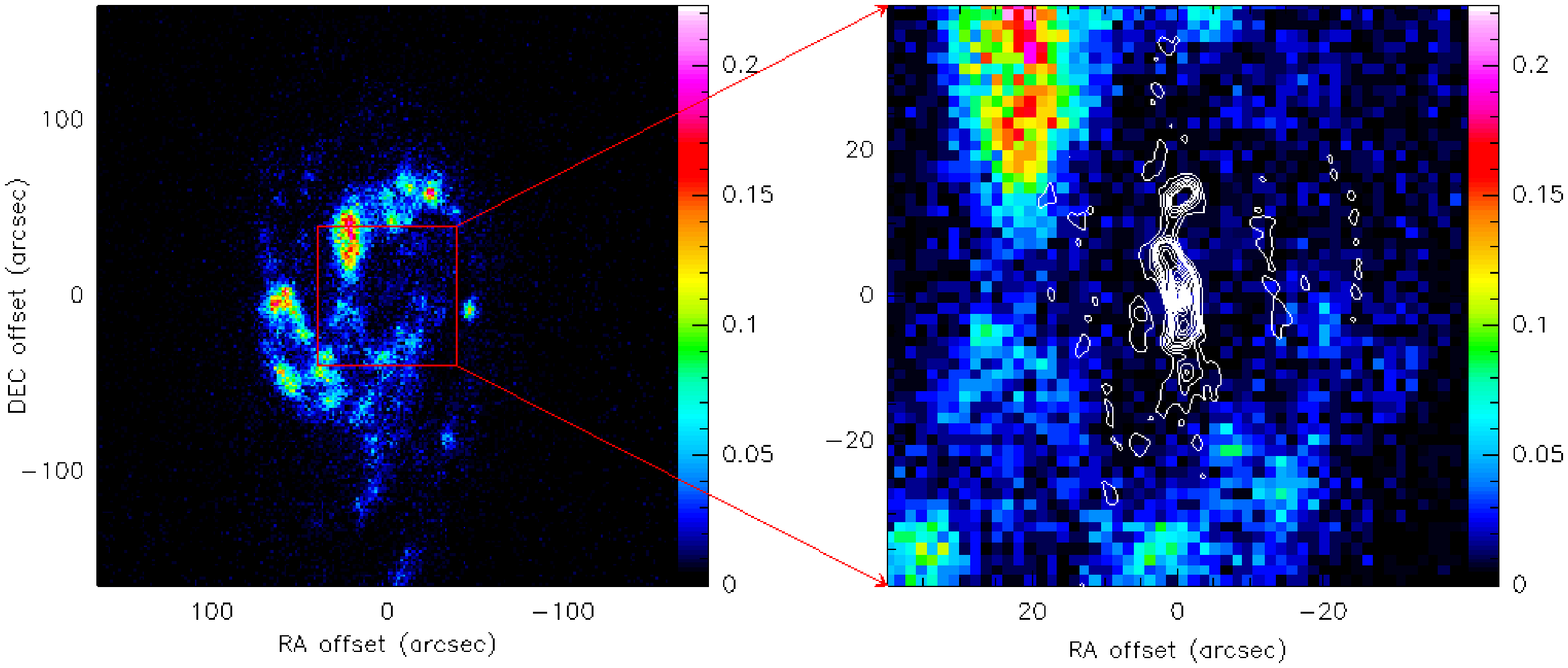}
\caption{\textit{Left panel}:  
Large-scale \textit{GALEX} FUV sky subtracted image of \nnn\, centered on the phase 
tracking center of our $^{12}$CO interferometric observations. 
The inner $330^{\prime\prime}$ are shown.
The (red) box shows the central $80^{\prime\prime}$, displayed in the 
\textit{right panel}. 
\textit{Right panel}: 
$^{12}$CO(1--0) integrated intensity contours (in white) as in Fig. \ref{fig:co10-21} 
(\textit{left panel}) overlaid on the \textit{GALEX} FUV sky subtracted image of \nnn. 
The inner $80^{\prime\prime}$ are shown.
}
\label{fig:galex_fuv-co}
\end{figure*}  

Another $H$-band image of \nnn\ is available thanks to the F160W filter on the 
\hst/NIC3 camera, described in Sect. \ref{sec:otherdata}.
Although the smaller FOV of the NIC3 camera (51\arcsec$\times$51\arcsec), 
the superior spatial resolution of the F160W/\hst\ $H$-band image (0\farcs2),
compared with that of the $H$-band 2MASS image (2\farcs5), allows to visualize more in 
detail the nuclear region of \nnn.
The right panel of Fig. \ref{fig:h-band} shows the inner 9\arcsec
of \nnn\ of the F160W/\hst\ $H$-band image with overlaid  $^{12}$CO(2--1) 
contours (in  green) as in Fig. \ref{fig:co10-21} (left panel). 
Although the morphology of the nuclear $^{12}$CO(2--1) emission is more elongated 
relative to the 1.6\,$\mu$m (F160W/\hst) one, the sizes of the two inner 
($\sim$3\arcsec) distributions are similar and the two central 
peaks coincide.
The central emission at 1.6\,$\mu$m is distributed in a disk,
with a central peak and a depression toward the NE caused by dust extinction. 
The dust feature gives a clue about the orientation of the galactic disk.
Because dust on the near side of the galaxy obscures what is behind it, we
can deduce that the eastern side is closer than the western side. 
Such an orientation would be consistent with the blue-shifted velocities toward
the north (see Fig. \ref{fig:co-vel}), and the expectation that
the spiral arms should be trailing.

\subsection{Far-infrared emission \label{sec:fir}}
Most of the dust mass in galaxies is relatively cool, 
radiating primarily at wavelengths of the \spitzer-MIPS 
70 and 160\,$\mu$m bands.
The left panels of Figures \ref{fig:mips70-bima} and \ref{fig:mips160-bima}, which 
display the inner 420\arcsec of the \spitzer-MIPS 70 and 160\,$\mu$m images
of \nnn, respectively, show that the dust emission is intensified at the nucleus and 
at the ansae at the ends of the bar.
The right panels of these figures display the BIMA $^{12}$CO(1--0) 
integrated intensity contours overlaid on the \spitzer-MIPS 70\,$\mu$m image 
(Fig. \ref{fig:mips70-bima}) and the \spitzer-MIPS 160\,$\mu$m image 
(Fig. \ref{fig:mips160-bima}) for the inner 200\arcsec\ of \nnn.
These comparisons show a very good correlation between the dust emission 
peaks and the $^{12}$CO emission along the bar.
These regions with strong dust and  $^{12}$CO emission could be the 
``highly obscured star-forming regions'' identified by \citet{prescott07} by
comparing 24\,$\mu$m and \ha\ maps. 
This dust-CO correlation is particularly important in terms of SF, 
discussed below in Sect. \ref{sec:sf}. 

\subsection{Far-ultraviolet morphology \label{sec:fuv}}
\begin{figure}
\includegraphics[angle=-90,width=0.45\textwidth]{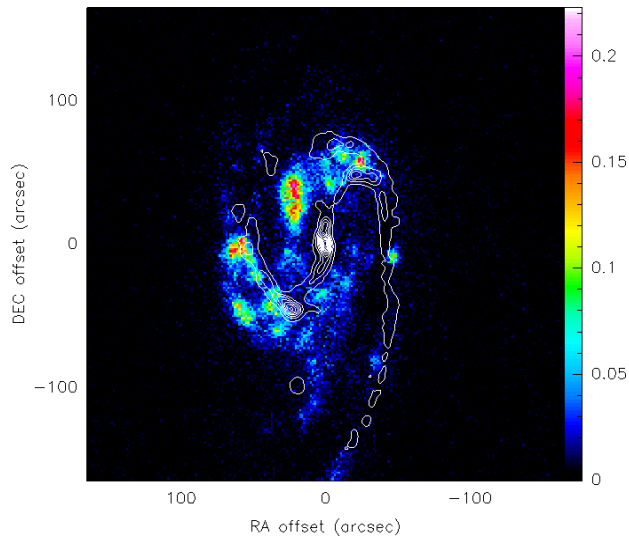}  
\caption{
BIMA $^{12}$CO(1--0) integrated intensity contours (in white) overlaid on the 
\textit{GALEX} FUV sky subtracted image of \nnn. 
The inner $320^{\prime\prime}$ are shown.
}
\label{fig:galex-bima}   
\end{figure}

The left panel of Figure \ref{fig:galex_fuv-co}  shows the 
FUV image of \nnn\ obtained with the \textit{GALEX} satellite, described in 
Sect. \ref{sec:otherdata}. 
It can be seen that \nnn\ has an asymmetric spiral 
structure in the FUV outside the range of our $^{12}$CO observations, 
as seen at other wavelengths (IR, but also $^{12}$CO from 
BIMA SONG survey).
There is an inner elongated (north/south) ring delimiting a 
``hole'', or rather a net depression in the FUV emission around the nucleus.  
Along the edge of this ring, the FUV emission is not homogeneously 
distributed, but its north/northeastern arc exhibits FUV emission 
in the form of several clumps stronger than the rest of the ring.
In the FUV, the eastern spiral arm, less extended,  appears 
more luminous than the western one, while
in the IR (8\,$\mu$m [Fig. \ref{fig:8mu}] and 1.6\,$\mu$m 
[Fig. \ref{fig:h-band}]) the two spiral arms are equally intense. 
The implication is that dust is suppressing the emission of the western arm.
A similar ring is visible also in \hi\ \citep[][]{haan08,haan09}, 
larger than in the FUV \textit{GALEX} 
emission but with the same elongated (north/south) morphology.
The right panel of Fig. \ref{fig:galex_fuv-co} shows our $^{12}$CO(1--0) 
contours as in Fig. \ref{fig:co10-21} (left panel) superimposed on 
the FUV \textit{GALEX} image for the inner 80\arcsec of \nnn.
The $^{12}$CO bar-like structure is contained in the inner hole observed 
in the FUV, indicating an anti-correlation between $^{12}$CO and FUV.
The same anti-correlation is observable by comparing the 
FUV \textit{GALEX} image with BIMA $^{12}$CO(1--0) contours, as 
displayed in Figure \ref{fig:galex-bima}.
Although the large scale views of the galaxy in FUV and $^{12}$CO(1--0) are 
similar (as demonstrated by the extension of the arms, especially 
the western one), in Fig. \ref{fig:galex-bima} we can appreciate some 
offsets between the two emissions.
Shifts between molecular gas and FUV peaks are present along both 
spiral arms, and the strongest FUV peak, along the ring in the north/northeast 
direction, has no counterpart in molecular gas emission.     

Offsets and anti-correlations between FUV and $^{12}$CO (or FIR and \hi) 
emissions have also been seen in other spiral galaxies such as M\,100 
\citep[][]{rand95,sempere97,calzetti05}, M\,51 \citep[][]{calzetti05}, 
and NGC\,3147 \citep[][]{vivi08a}.
These anti-correlations may relate to star formation efficiency and 
timescale variations in response to a spiral density wave.
The SF and its related tracers are often located in different 
regions of a galaxy: FUV emission is more prominent at the outer edge 
of the spiral arms, where typically dust extinction is low, 
while FIR emission is stronger at the inner edge.

The FUV-CO anti-correlation found in \nnn\ is particularly interesting 
in terms of the Kennicutt-Schmidt (KS) law \citep[][]{kennicutt98}, 
which relates the star formation rate (SFR) density ($\Sigma_{\rm SFR}$) 
to the gas surface density ($\Sigma_{\rm gas}$).
\citet{bigiel08} have found that at high gas surface densities, 
when the gas is predominantly molecular, the KS law is linear 
($\Sigma_{\rm SFR}$$\sim$$\Sigma_{\rm gas}^N$, with $N$$\sim$1) 
and a correlation between FUV and CO is expected. 
However, more than one factor may disturb the KS correlation,
especially on small spatial scales. First, FUV traces older SF episodes
than either \ha\ or 24\,\micron\ emission; hence the sites of
potential future SF (as traced by CO) may be disconnected
from past SF sites (FUV).
Second, feedback from massive stellar winds and supernovae may disrupt 
the ISM on small spatial scales and degrade the KS correlation.
Finally, if the molecular gas is not in dynamical equilibrium, perhaps
through the action of the large-scale bar, it would not be expected 
to be associated with sites of current SF.
The lack of correlation in the context of the KS law for this and 
other NUGA galaxies will be discussed in a forthcoming paper
dedicated to this topic.

\subsection{Star formation in NGC\,3627\label{sec:sf}}
The FIR luminosity is often used as a measure of the current star 
formation rate (SFR), since it is assumed that FIR emission is mainly 
due to dust heating by massive young stars.
The total IR luminosity of \nnn\ is 1.3$\times10^{44}$\,erg\,s$^{-1}$,
according to the precepts of \citet{draine07} and with fluxes from \citet{dale05}.
This corresponds to a SFR of $\sim6$\,M$_\odot$\,yr$^{-1}$ \citep{kennicutt98}.
In the bulge of \nnn, there is little observed SF \citep[][]{smith94,regan02}, and
the SFR given by \ha\ within a nuclear region of diameter $\sim$16\arcsec\ is
0.078\msun yr$^{-1}$ \citep[][]{regan02}, 
$\sim$3 times lower than
found in the bar itself, and $\sim$4 times lower than the spiral arms \citep[][]{regan02}.
Part of this deficit in the nuclear \ha-derived SFR may arise from dust extinction,
given that the mean A$_V$ in the inner 50\arcsec\ (diameter) is
$\sim$2\,mag \citep{calzetti07}.
In any case, in \nnn, most of the SF is extranuclear,
along the bar, particularly where it terminates and the spiral arms emerge
(see Figs. \ref{fig:mips70-bima} and \ref{fig:mips160-bima}).

The 70\,\micron\ emission is confined mainly to the nucleus and
the bar, particularly the ansae (see Fig. \ref{fig:mips70-bima}).
The 160\,\micron\ emission (albeit with lower resolution), 
is more broadly distributed, especially {\it around} the bar. 
This IR morphology suggests that the dust along the bar is warmer than 
around the bar, probably heated by the massive stars in the recent 
star-formation episodes.

In galaxies with weak SF activity \citep[e.g. NGC 4736,][]{smith94}, dust heating 
by non-OB stars may also contribute significantly \citep[e.g.,][]{dejong84,bothun89}.
This more quiescent heating source may be especially important in the central regions of 
early/type spiral galaxies with massive bulges and little nuclear or circumnuclear SF, 
such as \nnn.
The ratio of FIR to \ha\ luminosity for the bulge of \nnn\ is of $\sim$8100, 
significantly larger than for the star-forming regions in this galaxy, between 
$\sim$1000 and $\sim$2000.
The L(FIR)/L(\ha) ratio is also higher than can be accounted for by obscured 
SF with a normal initial mass function, using extinction measurements derived 
from $^{12}$CO(1--0) and FIR data \citep[][]{smith94}.
Thus, the older stars probably contribute significantly to the dust heating 
in the bulge of \nnn\ \citep[][]{smith94}.

A low nuclear SFR is consistent with the CO/HCN ratio (10) discussed in Sect. \ref{sec:30m}.
Higher ratios suggest that excitation by SF is dominant over AGN 
excitation in the circumnuclear region, but we found a ``normal'' CO/HCN ratio 
for \nnn, not surprisingly given its low SFR. 

\begin{figure}
\includegraphics[angle=-90,width=0.4\textwidth]{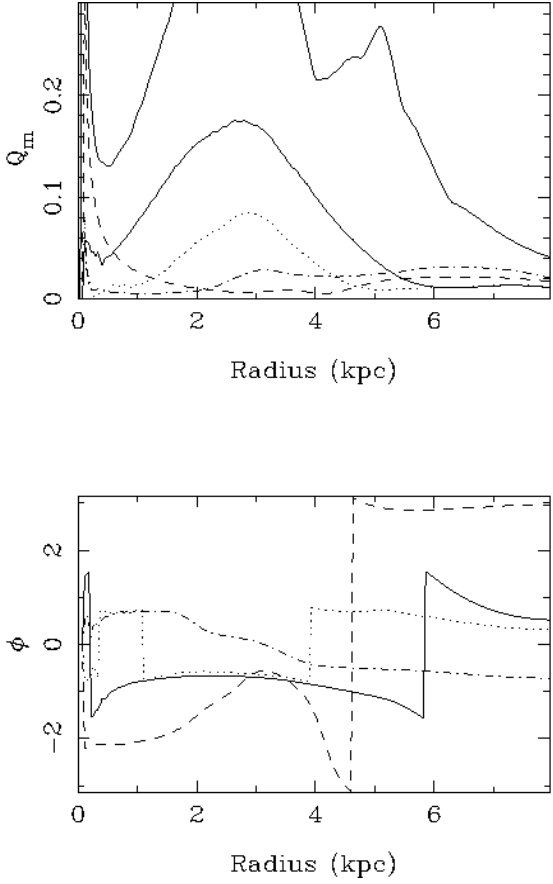}  
\caption{
The strength $Q$ (top) and phase $\Phi$ (bottom) of the $m$ = 1, 2, 3, 4 Fourier
components of the potential, derived from the \spitzer-IRAC 3.6\,\micron\ image.
The full lines correspond to $m=2$ and the total strength,
the dashed line to $m=1$, dot-dash to $m=3$, and dots to  $m=4$.
}
\label{fig:potn3627}   
\end{figure} 

\section{Computation of the torques on the molecular gas\label{sec:torques}}
The  gravitational torques derived from the stellar potential 
in the inner region of \nnn\ allow to account for the gas kinematics
derived from CO and examine the efficiency of gravitational torques 
exerted on the gas.
As described in previous NUGA papers \citep[e.g.,][]{santi05}, 
to compute the  gravitational torques we assume that NIR images give 
the best approximation for the total stellar mass distribution, 
being less affected than optical images by dust extinction or stellar 
population bias.

\subsection{Evaluation of the gravitational potential\label{sec:grav-pot}}
We computed the torques using both \hst-NICMOS F160W and 
\spitzer-IRAC 3.6\,\micron\ images.  
They yield complementary results, the torques computed from the 
\hst-NICMOS F160W image compared with the $^{12}$CO PdBI contours 
allow to investigate the nuclear region of \nnn, while the torques 
derived from the \spitzer-IRAC image in combination with the 
$^{12}$CO BIMA contours are much better adapted to visualize 
the whole spiral structure of the galaxy.
We perform the subtraction of foreground stars, deprojection, and resampling, 
as described in other NUGA papers \citep[e.g.,][]{santi05}.

\begin{figure}
\includegraphics[angle=-90,width=0.45\textwidth]{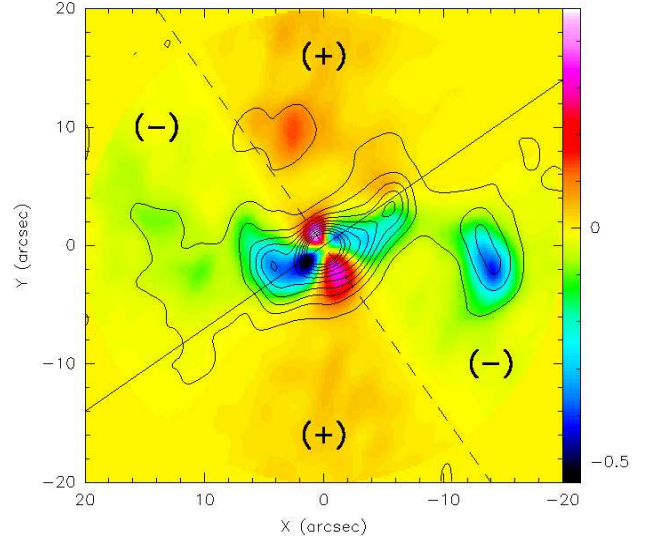}
\caption{
NUGA PdBI $^{12}$CO(1--0) contours are overlaid onto the gravitational 
torque map computed from the \hst-NICMOS F160W image 
(t(x,y)~$\times$~$\Sigma$(x,y), as defined in text) in the center of \nnn.
The map is deprojected, and rotated so that 
the major axis of the galaxy is oriented parallel to the abscissa Ox.
The continuum line is oriented along the large-scale bar in the plane 
of the galaxy (PA = 35$^\circ$), the dashed one orthogonally to the continuum 
line.
The quadrants labeled with (+) refer to positive  torques, while 
those labeled with (-) to negative torques.
The derived torques change sign as expected in a \textit{butterfly}
diagram.
}
\label{fig:n3627-torq1}
\end{figure}

Here, we briefly recall some definitions and assumptions used to evaluate 
the gravitational torques.
NIR images are completed in the vertical dimension by assuming an isothermal 
plane model with a constant scale height, equal to $\sim$1/12th of the radial 
scale-length of images.
With a Fourier transform method we derive the potential and we assume a 
constant mass-to-light (M/L) ratio able to reproduce the observed 
$^{12}$CO RCs. 
Beyond a radius of 20\arcsec (or 1.96\,kpc in diameter), the mass density is set to 
0 in the \hst-NICMOS F160W image, thus suppressing any spurious $m=4$ terms. 
This assumption is sufficient to compute the potential over the PdBI $^{12}$CO(1--0) 
primary beam.
For the \spitzer-IRAC 3.6\,\micron\ image, this radius truncation is done at 169\arcsec (or
16.6\,kpc in diameter).

For the non-axisymmetric part of the potential $\Phi(R,\theta)$, 
we decompose $\Phi(R,\theta)$ in Fourier components ($m$-modes), 
following \citet{francoise81}:
$$
\Phi(R,\theta) = \Phi_0(R) + \sum_{m=1}^\infty \Phi_m(R) \cos [m \theta - \phi_m(R)]
$$
\noindent
where $\Phi_m(R)$ and $\phi_m(R)$ are the amplitude and phase of the 
$m$-mode, respectively.

The strength of each $m$-Fourier component, $Q_m(R)$, is defined by the 
ratio between tangential and radial forces, 
$Q_m(R)=m \Phi_m / R | F_0(R) |$.
The strength of the total non-axisymmetric perturbation is defined by:
$$
Q_T(R) = {F_T^{max}(R) \over F_0(R)} 
$$
\noindent
where $F_T^{max}(R)$ and $F_{0}(R)$ represent the maximum amplitude of the 
tangential force and the mean axisymmetric radial force, respectively.
Figure \ref{fig:potn3627} shows the strengths (top panel) and phases 
(bottom panel) vs. radius for the \spitzer-IRAC 3.6\,\micron\ image and for the 
first $m$ components. 

\begin{figure}
\includegraphics[angle=-90,width=0.45\textwidth]{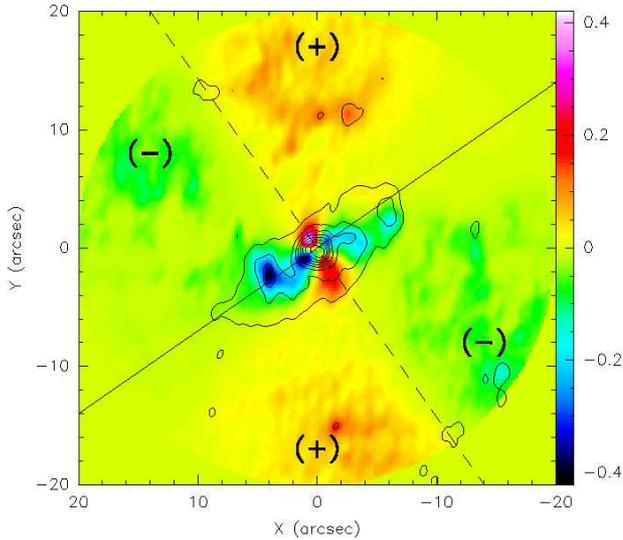}
\caption{
Same as Fig. \ref{fig:n3627-torq1} for our NUGA PdBI $^{12}$CO(2--1)
emission taken as tracer of gas surface density. 
}
\label{fig:n3627-torq2}
\end{figure}

\begin{figure}
\includegraphics[angle=-90,width=0.65\textwidth]{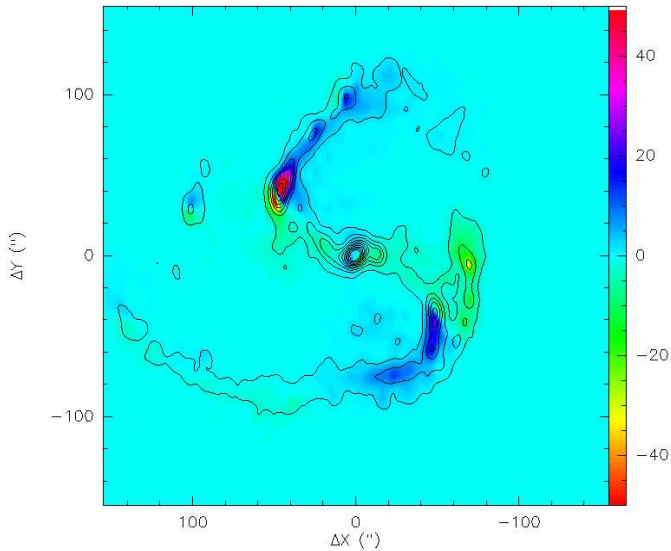}
\caption{
The BIMA $^{12}$CO(1--0) contours are overlaid onto the gravitational 
torque map computed from the \spitzer-IRAC 3.6\,\micron\ image.
As for Fig. \ref{fig:n3627-torq1}, the torque map (color scale) is 
plotted on a symmetric palette (wedge), and is deprojected and rotated 
so that the major axis of the galaxy is oriented parallel to the abscissa 
Ox.
}
\label{fig:n3627-torq1b}
\end{figure}

\subsection{Evaluation of the gravity torques\label{sec:grav-tor}}
The forces per unit mass ($F_x$ and $F_y$), obtained from the  
derivatives of $\Phi(R,\theta)$ on each pixel, allow to compute 
the torques per unit mass $t(x,y)$ by:
$$
t(x,y) = x~F_y -y~F_x.
$$
The torque map is oriented according to the sense of rotation 
in the plane of the galaxy.
The combination of the torque map and the gas density $\Sigma$ map 
allows to derive the net effect on the gas at each radius.
Figures \ref{fig:n3627-torq1} and \ref{fig:n3627-torq2} show 
gravitational torque maps, computed from the \hst-NICMOS F160W image,  
weighted by the gas surface density $t(x,y)\times \Sigma(x,y)$, 
normalized to their maximum value, for NUGA PdBI $^{12}$CO(1--0) and 
$^{12}$CO(2--1), respectively.
The difference in orientation between  
the large-scale bar and the major axis of the galaxy ($\sim$19$^\circ$ in the sky plane)
implies that the deprojected difference in PAs is $\sim$35$^\circ$. 
In both figures, the continuum line is oriented along the large-scale 
bar in the plane of the galaxy (PA = 35$^\circ$), the dashed one orthogonally 
to the continuum line.
In the two quadrants labeled with (+) the torques are positive, while 
in those labeled with (-) the torques are negative.
The derived torques change sign following a characteristic 2D 
\textit{butterfly} pattern produced by the bar. 
Figure \ref{fig:n3627-torq1b} shows the gravitational 
torque map, derived from the \spitzer-IRAC 3.6\,\micron\ image, 
weighted by the gas surface density $t(x,y)\times \Sigma(x,y)$, 
normalized to their maximum value, for BIMA $^{12}$CO(1--0).
The observed gas distribution is representative of the time 
spent by a molecular cloud on a typical orbit at this location. 

\begin{figure*}
\centering
\includegraphics[width=\textwidth,bb=46 145 775 392]{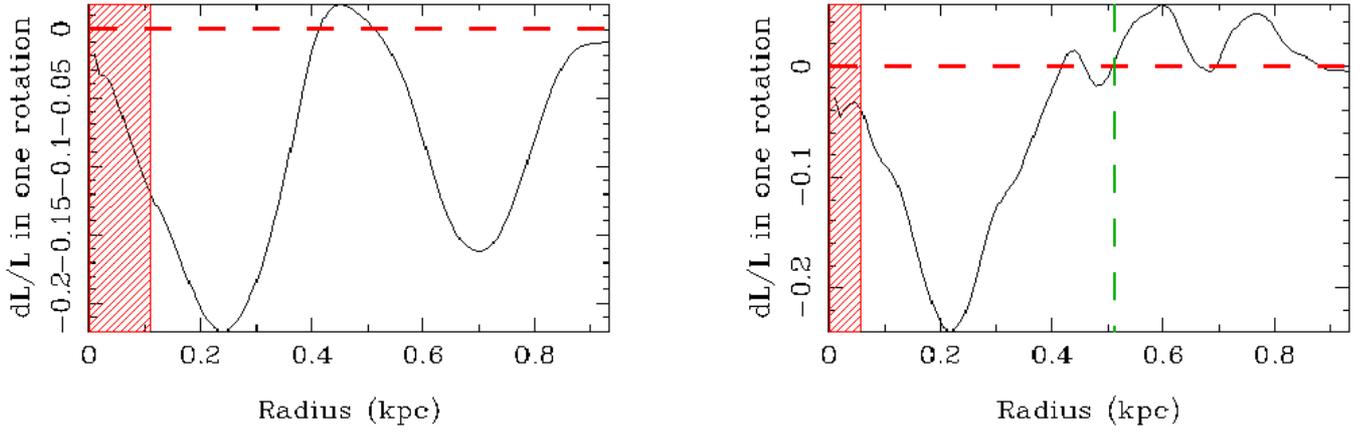}
\caption{
The torque, or more precisely the fraction of the 
angular momentum transferred from/to the gas in one rotation 
($dL/L$), computed with the \hst-NICMOS F160W image, is plotted for 
$^{12}$CO(1--0) [left] and $^{12}$CO(2--1) [right]. 
The (red) dashed area corresponds to the resolution limit 
of our observations. 
In the left panel, the limiting factor is the $^{12}$CO(1--0) map resolution 
($\sim$125\,pc), while, on the right, the $^{12}$CO(2--1) map resolution ($\sim$59\,pc). 
In the right panel, the (green) vertical dashed line at 0.51\,kpc indicates 
the $^{12}$CO(2--1) FOV at the PdBI.
}
\label{fig:gastor-nicmos}   
\end{figure*} 

\begin{figure*}
\centerline{
\hbox{
\includegraphics[width=0.4\textwidth,angle=-90]{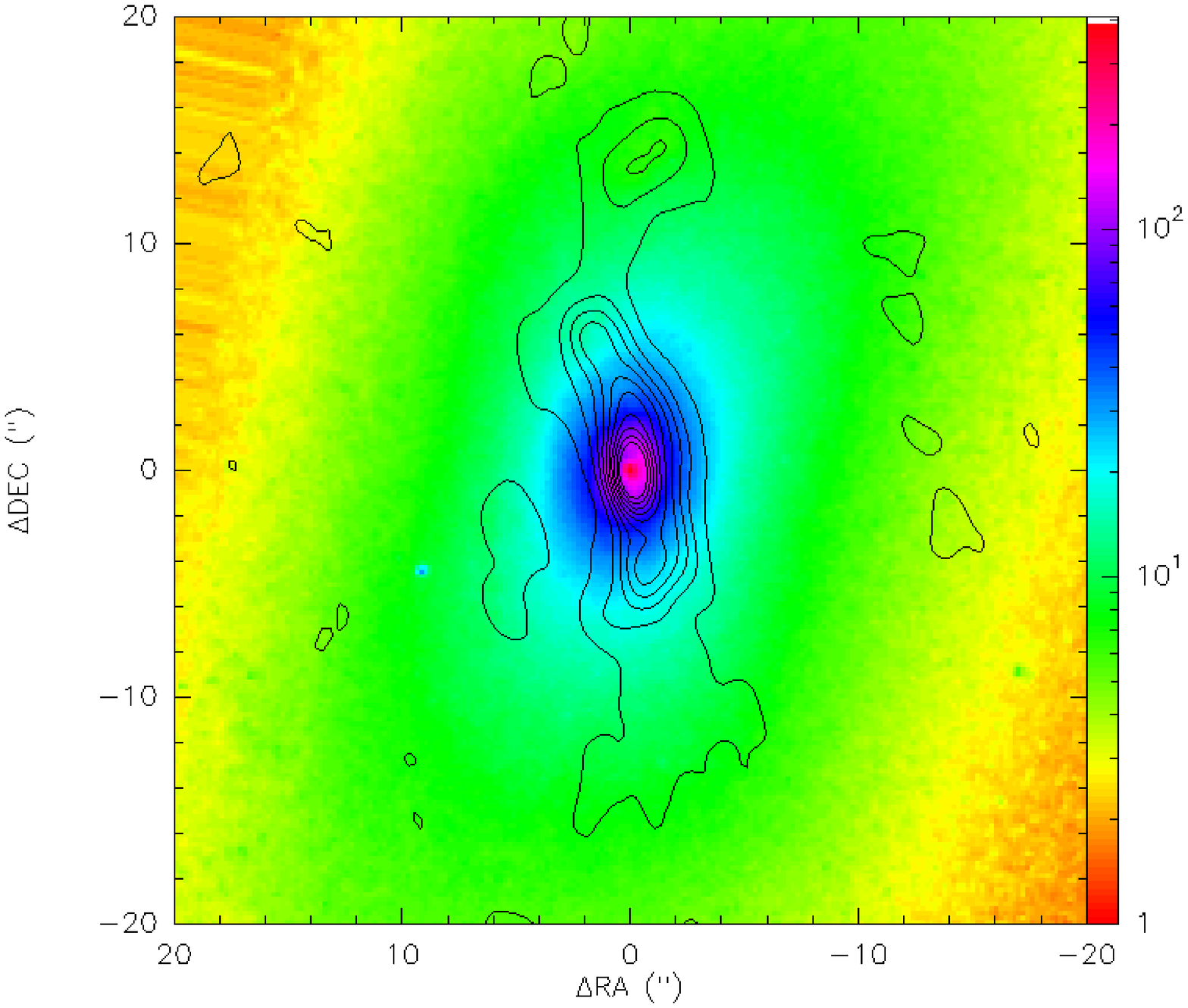}
\includegraphics[width=0.4\textwidth,angle=-90]{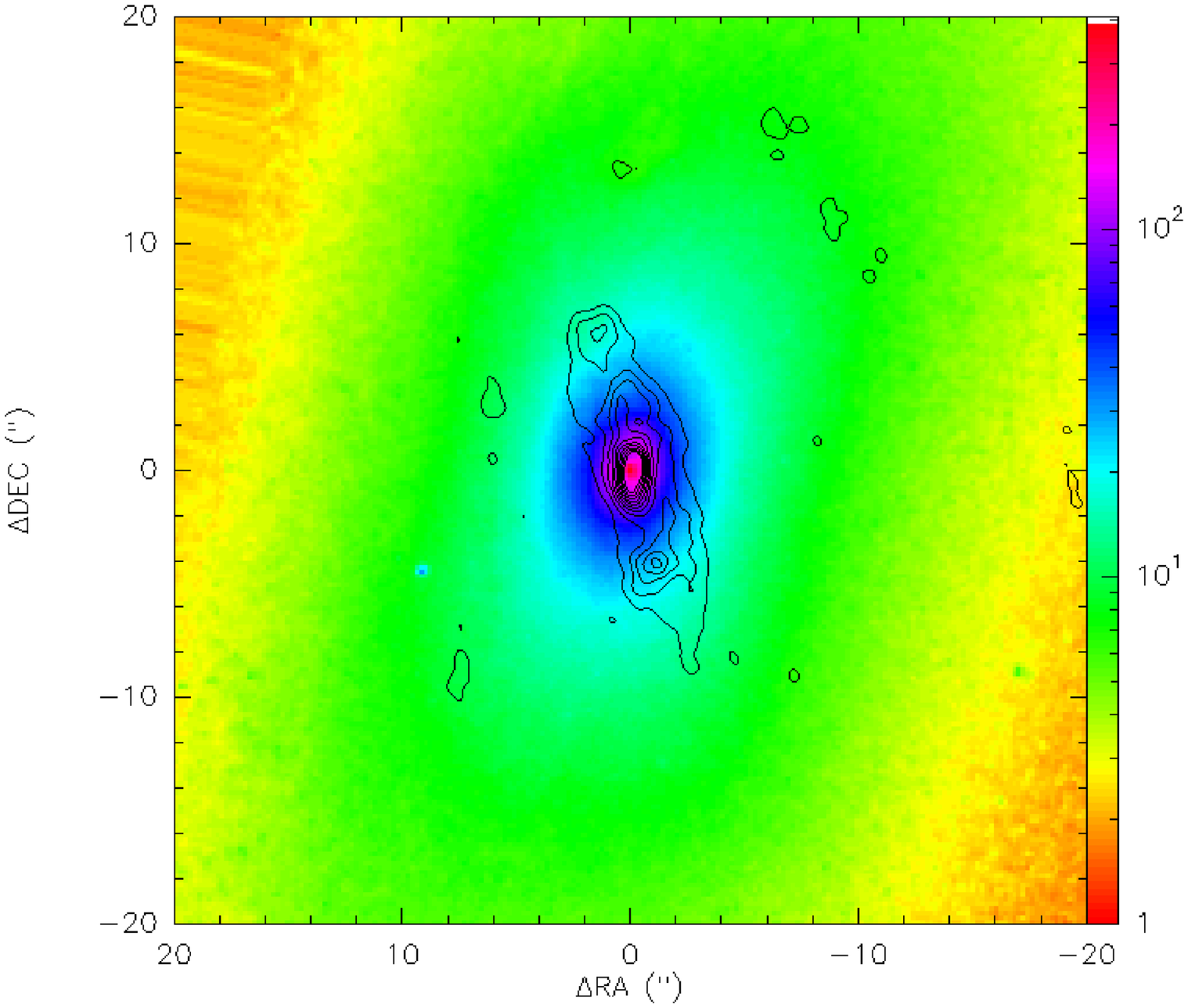}
}
}
\caption{
\textit{Left panel}: 
Our NUGA PdBI $^{12}$CO(1--0) contours overlaid on the \hst-NICMOS 
F160W image of \nnn\ in false color.
The inner 40\arcsec\ are shown.
\textit{Right panel}: 
Same as left panel but for the $^{12}$CO(2--1) line.
}
\label{fig:co-nicmos}
\end{figure*} 

By using $\Sigma(x,y)$ as the actual weighting function, we first 
compute the torque per unit mass averaged over azimuth:
$$
t(R) = \frac{\int_\theta \Sigma(x,y)\times(x~F_y -y~F_x)}{\int_\theta \Sigma(x,y)}
$$
where $t(R)$ is, for definition, the time derivative of the specific 
angular momentum $L$ of the gas averaged azimuthally, 
$t(R)$=$dL/dt~\vert_\theta$.   
Then, to have dimensionless  quantities, we normalize this variation 
of angular momentum per unit time to the angular momentum at 
this radius and to the rotation period.
Finally, we estimate the efficiency of the gas flow as the average 
fraction of the gas specific angular momentum transferred 
in one rotation ($T_{rot}$) by the stellar potential, as a 
function of radius:
$$
{\Delta L\over L}=\left.{dL\over dt}~\right\vert_\theta\times \left.{1\over L}~\right\vert_\theta\times 
T_{rot}={t(R)\over L_\theta}\times T_{rot}
$$
\noindent
where $L_\theta$ is assumed to be well represented by its axisymmetric 
estimate, $L_\theta=R\times v_{rot}$.
Figures \ref{fig:gastor-nicmos} show $\Delta L/L$ curves computed with 
the \hst-NICMOS F160W image for \nnn\ derived from the PdBI $^{12}$CO(1--0) 
[left] and the $^{12}$CO(2--1) [right] data. 
The (red) dashed area corresponds to the resolution limit of our observations. 
In the left panel, the $^{12}$CO(1--0) map resolution is $\sim$125\,pc, while 
in the right panel the $^{12}$CO(2--1) map resolution is $\sim$59\,pc.
In the right panel, the (green) vertical dashed line at 0.51\,kpc indicates 
the $^{12}$CO(2--1) FOV at the PdBI.
These figures show that the torques are negative within the inner 
0.4\,kpc, down to the resolution limit of our observations, and reach 
a (negative) peak at 0.2\,kpc for both $^{12}$CO(1--0) and $^{12}$CO(2--1).
For $^{12}$CO(1--0), they oscillate and become negative again at 700\,pc.
This second (negative) peak is not present in $^{12}$CO(2--1), because it is
outside the $^{12}$CO(2--1) FOV. 
Figures \ref{fig:co-nicmos} show NUGA PdBI $^{12}$CO(1--0) [left] 
and the $^{12}$CO(2--1) [right] contours overlaid on the \hst-NICMOS F160W image.
This figure clearly shows that the gas is leading the bar 
(the sense of the rotation is direct) since the PA of the bar is $-21^{\circ}$, 
while the bulk emission of the molecular gas is distributed along a bar-like 
structure with a PA of $14^{\circ}$ for the $^{12}$CO(1--0) line and of 
$15^{\circ}$ for the $^{12}$CO(2--1) one (see Fig. \ref{fig:co-vel} and 
Sect. \ref{sec:cokinematics}).   
This explains the negative torques observed in Figs. \ref{fig:gastor-nicmos}.  

\begin{figure}
\centering
\includegraphics[width=0.45\textwidth,bb=227 195 570 450]{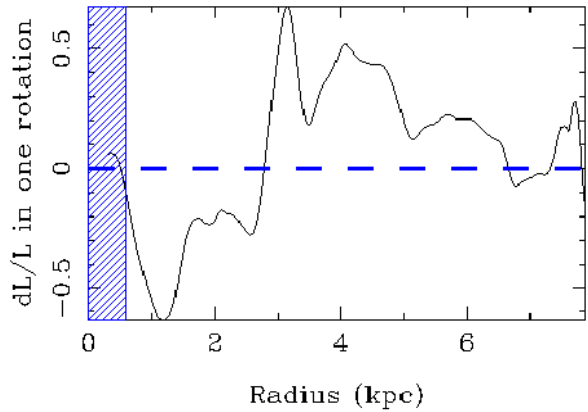}
\caption{
The torque, or more precisely the fraction of the 
angular momentum transferred from/to the gas in one rotation 
($dL/L$), computed with the \spitzer-IRAC 3.6\,\micron\ image, is 
plotted for BIMA $^{12}$CO(1--0) observations. 
The (blue) dashed area corresponds to the resolution limit of the BIMA 
observations ($\sim$670\,pc).  
}
\label{fig:gastor-irac}   
\end{figure}

Figure \ref{fig:gastor-irac} shows the large-scale $\Delta L/L$ curves computed with 
the \spitzer-IRAC 3.6\,\micron\ image and the BIMA-SONG images.
The (blue) dashed area corresponds to the resolution limit of BIMA observations, 
$\sim$670\,pc.   
In this figure, the torques are negative within the inner $\sim$3\,kpc, 
down to the resolution limit of BIMA maps, and positive at larger radii.
Resonances in barred galaxies are related to the pattern speed of the stellar bar, 
which can be inferred from the angular velocity at the corotation (CR) radius.
In early-type barred galaxies, CR is located near the end of the bar between 
$R=R_{\rm bar}$ and $R=1.4 R_{\rm bar}$ \citep[][]{elmegreen96}.
As said before, in \nnn, the torques are negative within the inner 
$\sim$3\,kpc, and the radius of $\sim$3\,kpc would correspond to 
the end of the bar, or to the CR radius estimated by \citet{chemin03} 
between 2.6 and 3.7\,kpc. 
The large-scale gas response inside the bar is twisted, i.e., 
the gas appears along the bar leading edges (see Fig. \ref{fig:co10-21}).
This is due to the dissipation there where the torques act, 
and it is a scenario which does not require any inner 
Lindblad resonance (ILR).
In any case, the PdBI-based torques from the NICMOS image
(Fig. \ref{fig:gastor-nicmos}) are more 
reliable in the center than the ones inferred from BIMA observations 
and the IRAC image (Fig. \ref{fig:gastor-irac}) 
because of the higher spatial resolution of the PdBI$+$NICMOS.

\section{Dynamical interpretation \label{sec:interpretation}}

\begin{figure}
\centering
\includegraphics[width=0.45\textwidth]{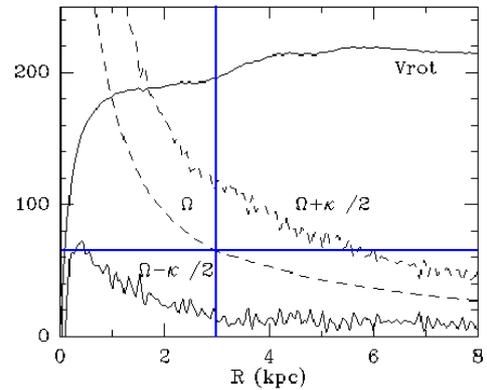}
\caption{
Model rotation curve V$_{\rm rot}$ (in km s$^{-1}$) and derived 
frequencies $\Omega$, $\Omega - k/2$, and $\Omega + k/2$ 
(in km s$^{-1}$ kpc$^{-1}$) for \nnn.  
The (blue) vertical line indicates the stellar bar end ($\sim$3\,kpc), while the 
(blue) horizontal one indicates the pattern speed of the bar 
at $\Omega_{\rm p}$$\sim$65\,km s$^{-1}$ kpc$^{-1}$. 
}
\label{fig:vcir}   
\end{figure}

Figure \ref{fig:vcir} shows the model RC (V$_{\rm rot}$) obtained from 
the \spitzer-IRAC 3.6\,\micron\ image, and the derived frequencies 
$\Omega$, $\Omega - k/2$, and $\Omega + k/2$ for \nnn.
We have also derived a model RC from the \hst-NICMOS F160W image, 
which is identical to the model RC from the \spitzer-IRAC 3.6\,\micron\ image 
until 1.3\,kpc from the center (it is not shown in the figure 
because it adds no new information).  
The model RC has been calibrated to the \ha\ RC derived from \citet{chemin03}, 
to have V = 212\,km s$^{-1}$ at  8\,kpc. 
Fig. \ref{fig:vcir} shows that if the stellar bar ends at $\sim$3\,kpc 
(blue vertical line), with a pattern speed of the bar 
$\Omega_{\rm p}$$\sim$65\,km s$^{-1}$ kpc$^{-1}$ (blue horizontal line)
then the CR of the bar would be at $\sim$3.3\,kpc, consistent with the 
results of \citet{chemin03} who estimated the CR radius between 
2.6 and 3.7\,kpc.
The $\Omega - k/2$ curve could be marginally compatible with an ILR 
at r = 500\,pc, but because there is no clear ring signature in the morphology,
we exclude the presence of an ILR.
In addition, since there is a strong bar, true precessing rate is lower 
than the epicyclic approximation $\Omega - k/2$ curve, and therefore 
the curve is only an upper limit. 
Hence there is probably no ILR.
The absence of an ILR was already discussed in Sect. \ref{sec:grav-tor}, 
where we elaborated such a scenario. 

The pattern speed of \nnn\ has been measured by several authors.
\citet{chemin03} have assumed that the inner ring-like feature found by 
\citet{regan02} but not by us, corresponds to the location of an 
ultra-harmonic resonance.
\citet{rand04}, using the BIMA SONG data and the Tremaine-Weinberg
method, have measured a pattern speed of the bar of 
$\Omega_{\rm p}=$50$^{+3}_{-8}$\,km s$^{-1}$ kpc$^{-1}$, 
in good agreement with our determination.

In the presence of an ILR, 
the torques should be negative between the CR of the bar and the ILR.
For \nnn, without an ILR, the torques are negative between the CR of 
the bar and the AGN, down to the resolution limit of our observations.
Such a conclusion would be the natural outcome of a
scenario in which a young and rapidly rotating bar has had no 
time to slow down due to secular evolution, and has not yet 
formed any ILR.
The presence of molecular gas inside the ILR of the primary bar, 
or where we expect that the ILR will form, makes \nnn\ a potential 
\textit{smoking gun} of inner gas inflow. 
In this scenario, the gas there is certainly fueling the central region, 
and in a second step could fuel directly the AGN.
Finding \textit{smoking gun} evidence of AGN fueling is 
proving to be quite challenging, perhaps because
of the short-lived nature of the mechanisms responsible.

\section{Summary and conclusions \label{sec:conclusions}}
The molecular gas, traced by $^{12}$CO(1--0) and $^{12}$CO(2--1)
transitions, in the interacting barred LINER/Seyfert 2 galaxy 
\nnn\ is distributed along a bar-like structure of 
$\sim$18\arcsec\ ($\sim$900\,pc) diameter (PA = 14$^{\circ}$) 
with two peaks at the extremes.
The 1.6\,$\mu$m $H$-band 2MASS and 3.6\,$\mu$m \spitzer-IRAC 
images of \nnn\ show a stellar bar in the nucleus with 
PA = $-21^{\circ}$, different from the PA (= $14^{\circ}$) 
of the molecular gas bar-like structure, suggesting that 
the gas is leading the stellar bar.
Instead, the \textit{GALEX} FUV emission of \nnn\ displays 
an inner elongated (north/south) ring delimiting a hole around the nucleus
and containing the $^{12}$CO $\sim$18\arcsec\ bar-like structure.
This kind of anti-correlation between FUV, molecular gas,
and the stellar bar is perhaps related to star formation efficiency 
and timescale variations in response to a spiral density wave.
The gravity torques exerted by the stellar potential on the gas
computed with the \hst-NICMOS F160W image and our 
PdBI maps are negative within the inner 0.4\,kpc, 
down to the resolution limit of our observations.
 
The torques computed with the \spitzer-IRAC 3.6\,\micron\ image and BIMA
$^{12}$CO map (with a resolution limit of $\sim$670\,pc) are also negative within 
the inner $\sim$3\,kpc.
If the bar ends at $\sim$3\,kpc, with a pattern speed of the bar 
$\Omega_{\rm p}$$\sim$65\,km s$^{-1}$ kpc$^{-1}$ then the CR of 
the bar would be at $\sim$3.3\,kpc.
There is no clear ring signature and we thus exclude the presence of an ILR.
Without an ILR, the torques are negative between the CR of 
the bar and the AGN, down to the resolution limit of our observations.
This scenario is compatible with a young/incipient bar which had no 
time to slow down from secular evolution, and has not yet formed any ILR.
\nnn\ is a potential \textit{smoking gun} of inner gas inflow:
the gas is certainly fueling the central region, 
and in a second step could fuel directly the AGN. 

\nnn\ is the fifth \textit{smoking gun} NUGA galaxy, together with 
NGC\,6574 \citep{lindt-krieg08},
NGC\,2782 \citep{leslie08},
NGC\,3147 \citep{vivi08a}, and
NGC\,4579 \citep{santi09}.
The common feature shared by these galaxies is a slowly rotating stellar
bar (or oval) with overlapping dynamical resonances \citep{vivi08b} associated with
kinematically decoupled inner bars or ovals.
Such resonances and kinematic decoupling are fostered by a large 
central mass concentration and high gas fraction.
\nnn\ is the unique potential \textit{smoking gun} NUGA galaxy 
with only one slowly rotating stellar bar. 
For \nnn, this drives a molecular bar-like structure, apparently
sufficient to transport the gas toward the AGN that, in a second step, 
could fuel directly the active nucleus. 

\begin{acknowledgements}
The authors would like to thank the referee, Greg Bothun, 
whose comments have been very useful for improving the original version 
of the paper.
We thank the scientific and technical staff at IRAM for their 
work in making our 30\,m and PdBI observations possible.
This research has made use of the NASA/IPAC Extragalactic Database (NED),
HyperLeda Database, {\it IRAS} Catalog, \spitzer\ archive, and
$MAST/GALEX$ images.
\end{acknowledgements}


\end{document}